\documentclass[aps,prb,notitlepage,superscriptaddress,twocolumn ]{revtex4-2}
\usepackage{graphicx, subfigure,epsfig}
 \usepackage{array,multirow}
\usepackage{dcolumn}
\usepackage{amssymb,amsmath,amsfonts,mathrsfs}
\usepackage{array}
\usepackage{times,setspace}
\usepackage{latexsym}
\usepackage{float,flafter, bm,bbm}
\usepackage{epstopdf, color,multirow}
\usepackage[colorlinks,linkcolor=blue,anchorcolor=blue,urlcolor=blue,citecolor=blue]{hyperref}
\usepackage{footnote}
\usepackage[T1]{fontenc}
\usepackage[normalem]{ulem}
\usepackage{xcolor}

\hypersetup{
    colorlinks=true,
    linkcolor=blue,
    filecolor=magenta,
    urlcolor=blue,
}

\begin{document}

\title{Mean Field Study of Superconductivity in the Square Lattice $t$-$J$ Model with Three-Site Hopping}

\author{Ke Yang}
\affiliation{Kavli Institute for Theoretical Sciences, University of Chinese Academy of Sciences, Beijing 100190, China}
\author{Qianqian Chen}
\affiliation{Kavli Institute for Theoretical Sciences, University of Chinese Academy of Sciences, Beijing 100190, China}
\author{Lei Qiao}
\affiliation{Kavli Institute for Theoretical Sciences, University of Chinese Academy of Sciences, Beijing 100190, China}
\author{Zheng Zhu}
\email{zhuzheng@ucas.ac.cn}
\affiliation{Kavli Institute for Theoretical Sciences, University of Chinese Academy of Sciences, Beijing 100190, China}
 \affiliation{CAS Center for Excellence in Topological Quantum Computation, University of Chinese Academy of Sciences, Beijing, 100190, China}

\date{\today}

\begin{abstract}
It remains an open question whether the two-dimensional single-band pure Hubbard model and its related pure $t$-$J$ model truly capture the superconducting order in cuprates.  Recent numerical studies on this issue have raised a notable disparity in superconducting order between the pure Hubbard model and the pure $t$-$J$ model. Inspired by these, we investigate the role of the three-site hopping term in $d$-wave superconductivity, such a term is usually neglected in the effective Hamiltonian of the Hubbard model, though its amplitude is of the same order as the superexchange coupling $J$ in the $t$-$J$ model.
Our slave-boson mean-field solution demonstrates the suppression of $d$-wave superconducting order by incorporating the three-site hopping term, consistent with numerical observations by the density matrix renormalization group. This suppression could be understood as a result of competition between superexchange interaction and three-site hopping, the former favors $d$-wave pairing while the latter favors $s$-wave pairing. We also discussed its role in quasiparticle dispersion and boson-condensation temperature.
Our findings may offer an alternative understanding of the recent numerical contrasting findings in the strong coupling regime: the absent or weak superconductivity in the pure Hubbard model, while the robust superconductivity in the $t$-$J$ model without including the three-site hopping term.
\end{abstract}

\maketitle

\begin{figure}[tbp]
	\includegraphics[width=0.48\textwidth]{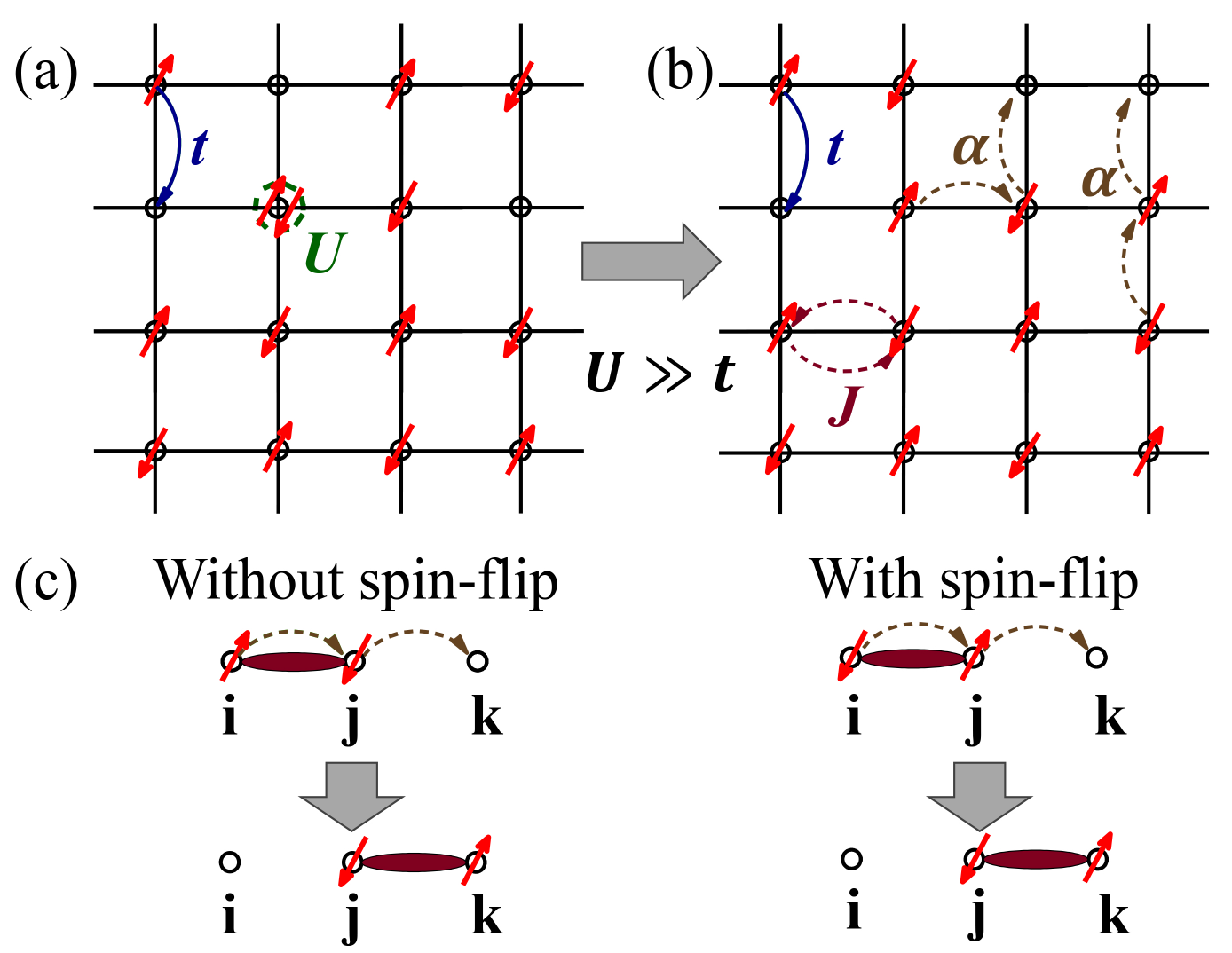}	
	\caption{\label{Figure1}(Color online) Graphical representation of the Hubbard model and its effective model. (a) The pure Hubbard model, where $t$ denotes the amplitude of the nearest-neighbor hopping, $U$ represents the  Coulomb repulsion. (b) The effective model obtained from the pure Hubbard model in the strong coupling limit is known as the $t$-$J$-$\alpha$ model. $J$ describes the superexchange interaction (blue dashed line); $\alpha$ represents the amplitude of three-site hopping (green dashed line), which exhibits collinear and non-collinear hopping shapes. $J=4t^2/U$ and $\alpha=J/4$  are of the same order.  (c) Two types of three-site hopping processes with hopping amplitude $\alpha$.  The left panel depicts a second-neighbor hopping via an intermediate site without a spin flip, and the right panel shows a second-neighbor hopping with a spin flip. The three-site hopping can also be viewed as pair hopping of a spin singlet.
	}
\end{figure}

\section{Introduction}
Since the discovery of high-temperature superconductivity in cuprates, numerous theories have been developed to understand their correlation physics~\cite{Review1,Review2,Review3,Review4}. It is widely acknowledged that the essential physics of cuprates can be described by the three-band Hubbard model~\cite{Cup2}. Within the  Zhang-Rice (ZR) singlet paradigm~\cite{Cup3}, the low-energy effective model simplifies to a one-band $t$-$J$ type model~\cite{Cup3,Htre1,Htre2}. Despite the consensus that the correlation effects in copper-oxygen planes play a pivotal role in the emergence of $d$-wave superconductivity~\cite{SingleH1,SymD1,SymD2}, it remains an open question whether the two-dimensional (2D) pure Hubbard model and its related pure $t$-$J$ model truly capture the superconducting (SC) order in cuprates~\cite{Review3,Review5,Review6,Review7,Review8,IfH1}.

Recent numerical studies on this issue have raised a notable disparity in SC order in the strong coupling regime between the pure Hubbard model and the pure $t$-$J$ model. Specifically, numerical simulations utilizing density matrix renormalization group (DMRG) and quantum Monte Carlo (QMC) suggest the absence of SC order in the pure Hubbard model over a range of doping $\delta\approx0.1\sim 0.2$ and interaction strength $U/t\approx6\sim8$~\cite{Hubb1}. By contrast, more recent DMRG calculations of the pure $t$-$J$ model~\cite{TJ1} demonstrate the robust SC order in the underdoped region. Moreover, tremendous efforts have also been devoted to exploring the fundamental physics in the theoretical models and characterizing the properties of real materials, for instance, the impact of long-range hoppings or Coulomb interactions~\cite{NHub1, TJ000, TJ00, TJ1, TJ2, TJ3, TJ4, TJ5, Ht0, Ht5, exmf1, exmf2, exmf5,exmfn}, incorporating the superexchange interaction into the extended Hubbard Hamiltonian~\cite{exmf4, exmf9}, the spin-charge mutual statistic ~\cite{Ht6, Ht7, Ht8, Ht9}, the additional gradients in multiple orbitals~\cite{Ht1, Ht2, Ht3, Ht4}, the bosonic modes such as phonons~\cite{ BM3, BM5, BM6}, the fluctuation effects~\cite{exmf6, exmf7}, the lattice anisotropy~\cite{exmf3, exmf8}, the interlayer hoppings or Coulomb interactions~\cite{cub0, cub1, cub2}  and other various  factors~\cite{Review1, Review2, Review3, Review4}.

 Theoretically, the pure $t$-$J$ model is often adopted as the effective Hamiltonian of the pure Hubbard model in the strong coupling limit, and the three-site hopping term~\cite{Auerbach1} is neglected despite its amplitude $\alpha$ being of the same order as the superexchange coupling $J$, i.e., $\alpha=t^2/U=J/4$~\cite{Review1, Review2, Review3}. Earlier studies have investigated the effects of the three-site hopping term on SC states in two-dimensional (2D) $t$-$J$ type models~\cite{3s1, 3s2, 3s2ed, 3s9, 3s8, 3s11,  ap201,3s15}. A Variational Monte Carlo (VMC) study~\cite{3s8} shows the three-site hopping term becomes quantitatively important for doping $\delta\ge 0.1$ and favors $s$-wave state. Mean-field analyses~\cite{3s11, ap201} support this finding and provide insight into the superconducting mechanism in models incorporating this term. Notably, a renormalized mean-field theory (RMFT) study~\cite{ 3s3} reports that the three-site hopping term suppresses $d$-wave superconductivity { when $\delta>0.1$}.  Their theoretical results aligned fairly with the experimental results at the optimal doping and in the overdoped regime, offering insight into the upper critical doping concentration.  Moreover, in one-dimensional (1D) models that incorporate the three-site hopping term, the ground-state phase diagrams have been investigated by exact diagonalization~\cite{3s0, tja3} and DMRG~\cite{tja5}. Beyond the debates on ground states, the spectral properties are also examined for both 1D and 2D models with such a term~\cite{Ht4,3s6, 3s7, 3s14}. However, there remains a lack of comprehensive analysis of the three-site term in 2D $t$-$J$ type models by combining microscopic analytical methods and unbiased large-scale numerical simulations.

We denote the pure $t$-$J$ model with a three-site hopping term as the $t$-$J$-$\alpha$ model. Since the $t$-$J$-$\alpha$ model is inherently employed as the effective model of the pure Hubbard Hamiltonian, $\alpha$ is fixed. However, recent quantum simulators realized by loading ultracold atoms onto the optical lattices have become promising platforms for simulating the two-dimensional (2D) strongly correlated lattice models~\cite{tja1, tja2}.  By incorporating a tunable three-site hopping amplitude $\alpha$, the $t$-$J$-$\alpha$ model has been widely used to simulate a driven Hubbard system over a broad parameter space~\cite{tja4,tja6,tja7}. {On the other hand, the strength of $\alpha$ is under debate among different proposed effective Hamiltonians derived from particular cases of the three-band model and is influenced by the detailed hopping strengths, potential energies, and Coulomb interactions of the oxygen and copper sites~\cite{Htre2, three-3s-3, three-band-new}. Therefore, considering a tunable strength of the three-site term is a reasonable way to understand its effects.}

Motivated by recent developments and numerical debates, in this work, we analytically and numerically investigate the $d$-wave superconductivity in the $t$-$J$-$\alpha$ model, with a tunable amplitude $\alpha$ on the square lattice. Our primary focus is identifying the role of the three-site hopping in SC order. By employing the slave-boson mean-field (SBMF) analysis, we have revealed that increasing the three-site hopping amplitude suppresses the $d$-wave SC order. We also perform DMRG simulations of the $t$-$J$-$\alpha$ model and confirm the role of three-site hopping in  $d$-wave superconductivity. Notably, this finding aligns qualitatively with the previous RMFT results~\cite{3s3}. Furthermore, we examine the impact of increased $\alpha$ on the quasiparticle dispersion and the boson-condensation temperature, thereby enriching our understanding of $d$-wave superconductivity under the SBMF framework. Inspired by recent numerical discrepancies, we explore the impact of system size and aspect ratio of the square lattice on SC order.  Compared to the pure $t$-$J$ model, the effective Hamiltonian of the pure Hubbard model is more sensitive to system size and geometry. Lastly, we discuss the $s$-wave solution at large $\alpha$.

\begin{figure}[tbp]%
	\includegraphics[width=0.48\textwidth]{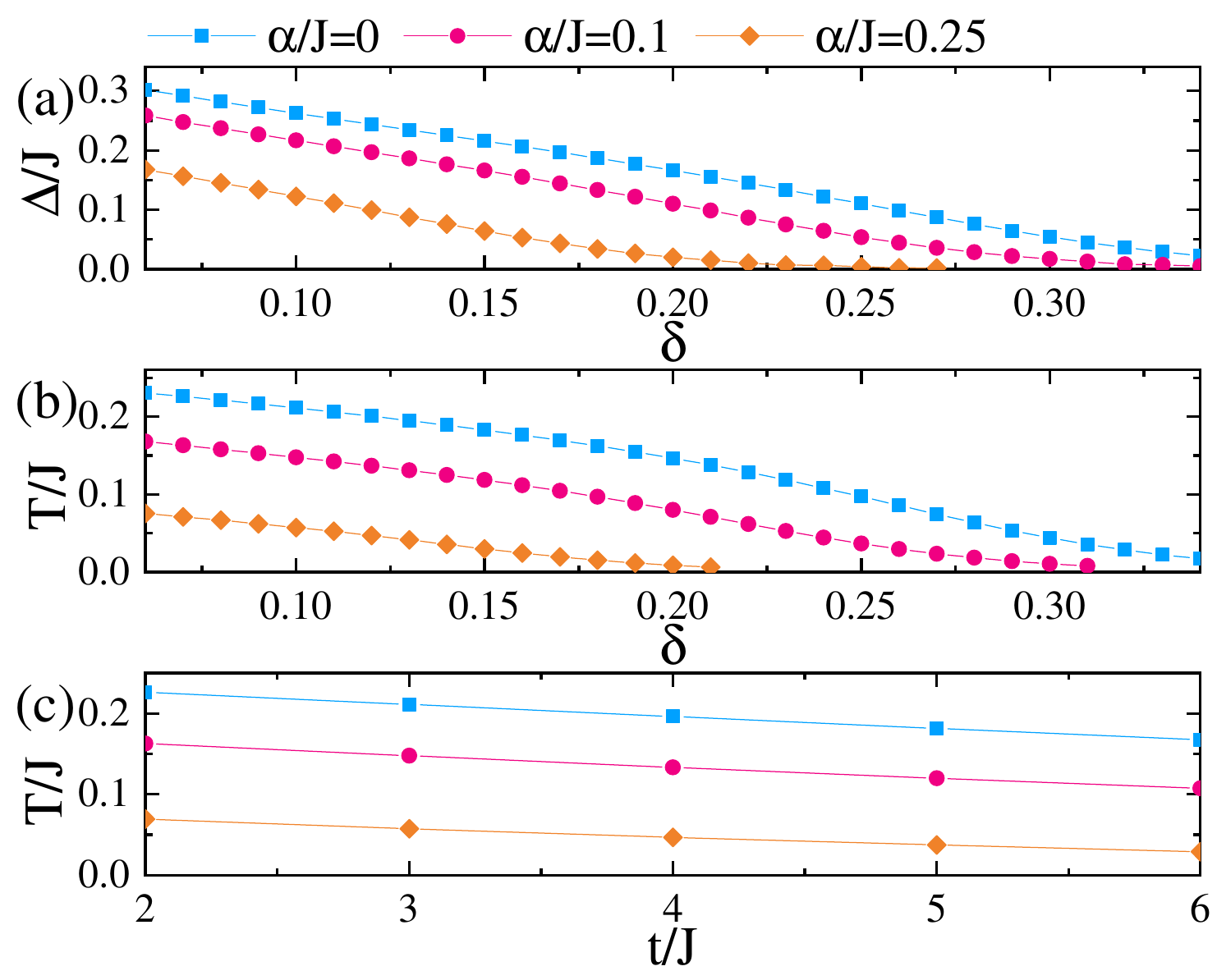}		
	\caption{\label{Figure2}(Color online)
Suppression of the $d$-wave superconductivity with the increase of three-site hopping amplitude $\alpha$. Here, we consider $\alpha/J$ $=0$ (square), $0.1$ (circle), and $0.25$ (diamond). Panels (a-b) show the RVB pairing order parameter $\Delta$ at $T=0$ (a) and the corresponding critical temperature $T_{\mathrm{RVB}}$ (b) as a function of doping concentration $\delta$ with fixed
$t/J=3$. At typical doping $\delta=0.1$, the critical temperature $T_{\mathrm{RVB}}$ as a function of coupling strength $t/J$ is illustrated in panel (c).
	}
\end{figure}

\begin{figure*}[tbp]%
	\includegraphics[width=0.43\textwidth]{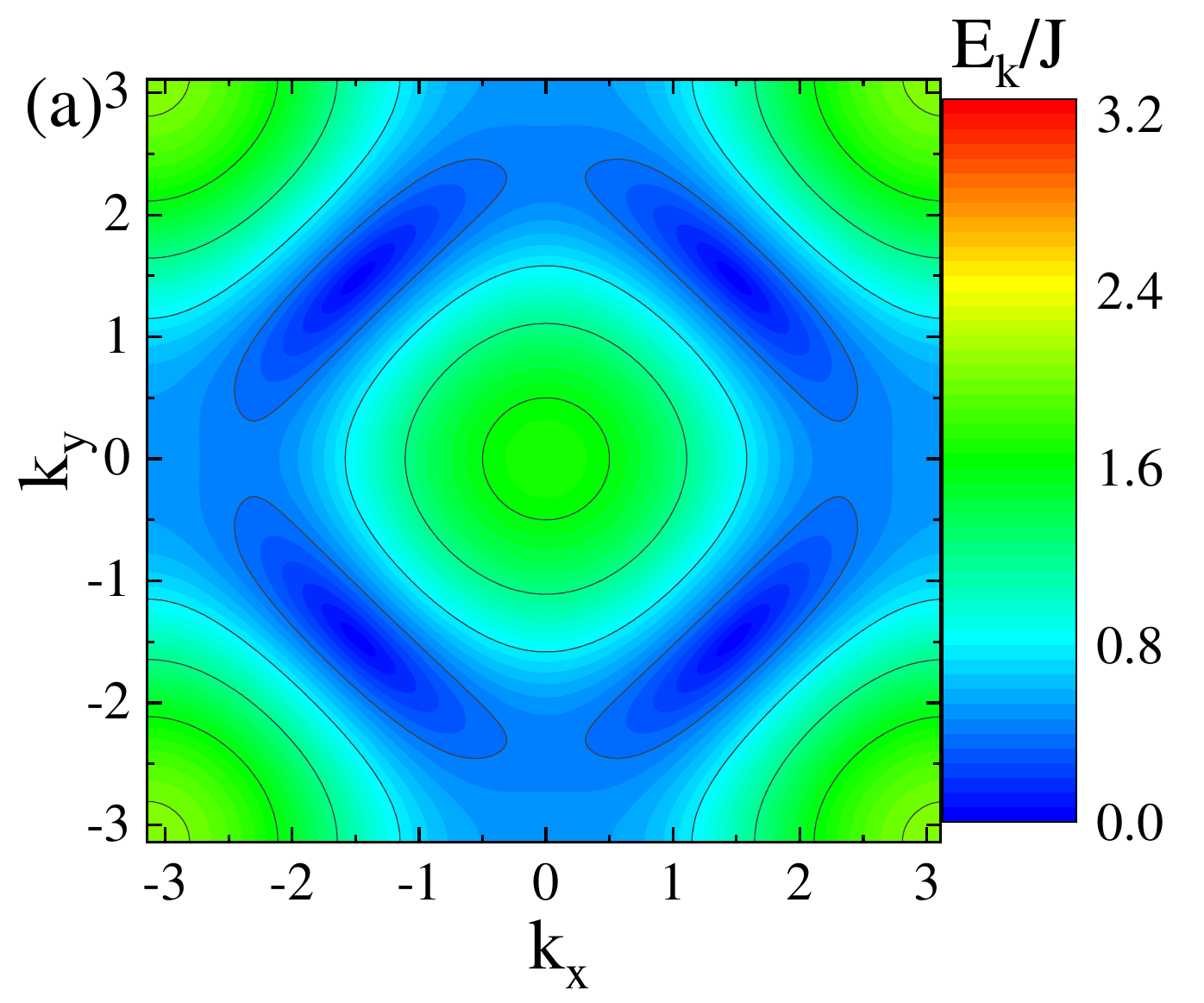}	
		\includegraphics[width=0.43\textwidth]{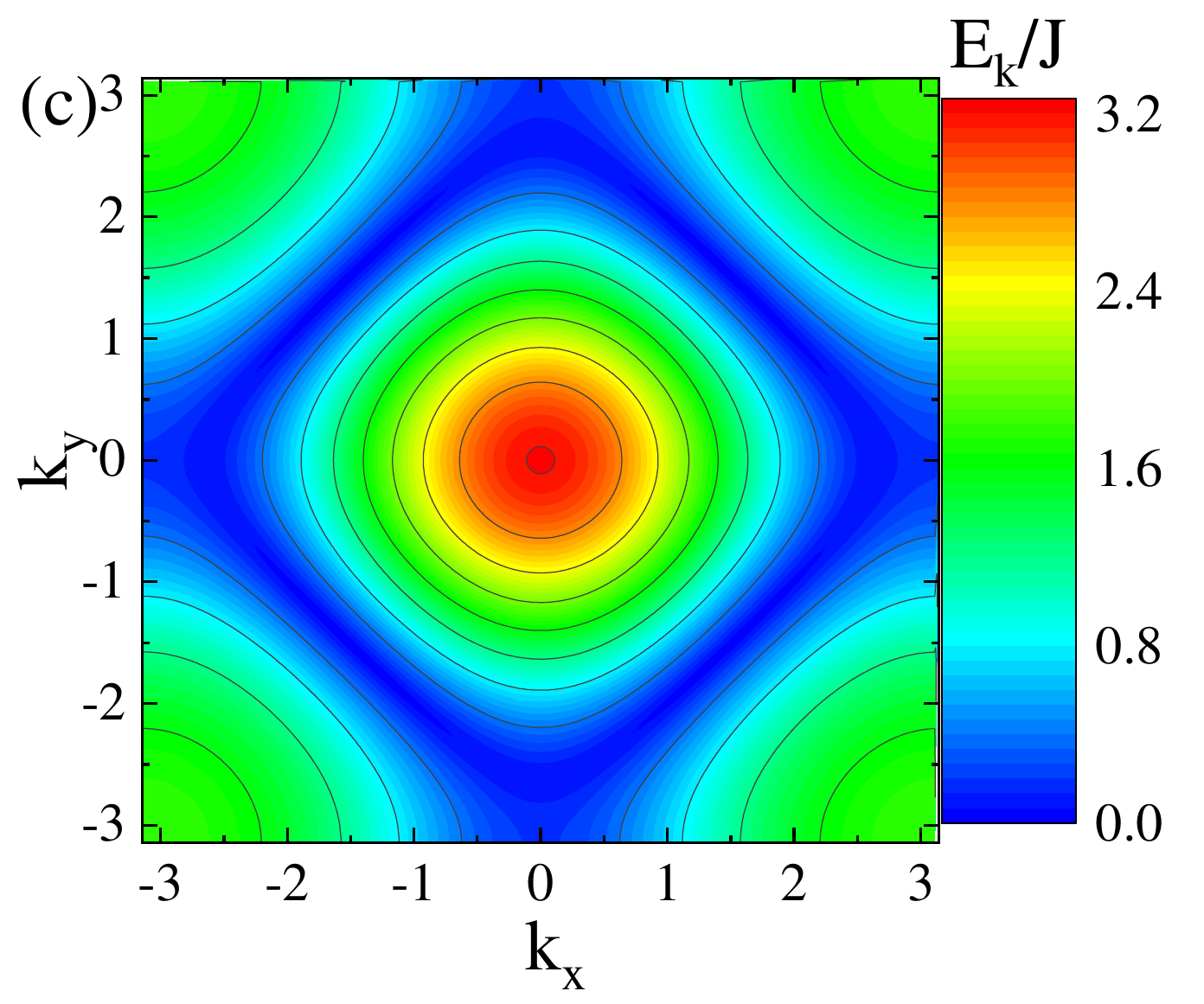}	
			\includegraphics[width=0.43\textwidth]{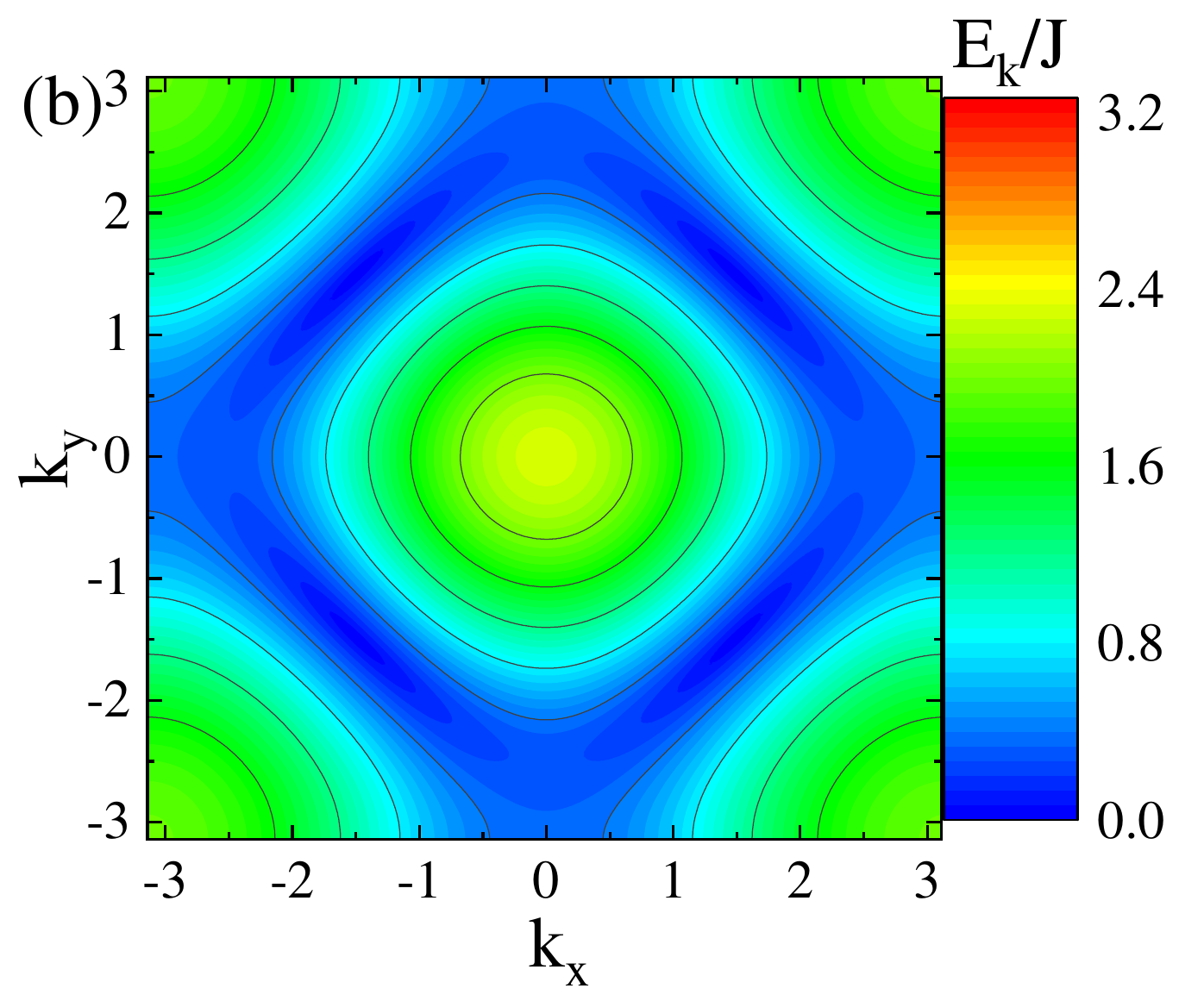}	
				\includegraphics[width=0.43\textwidth]{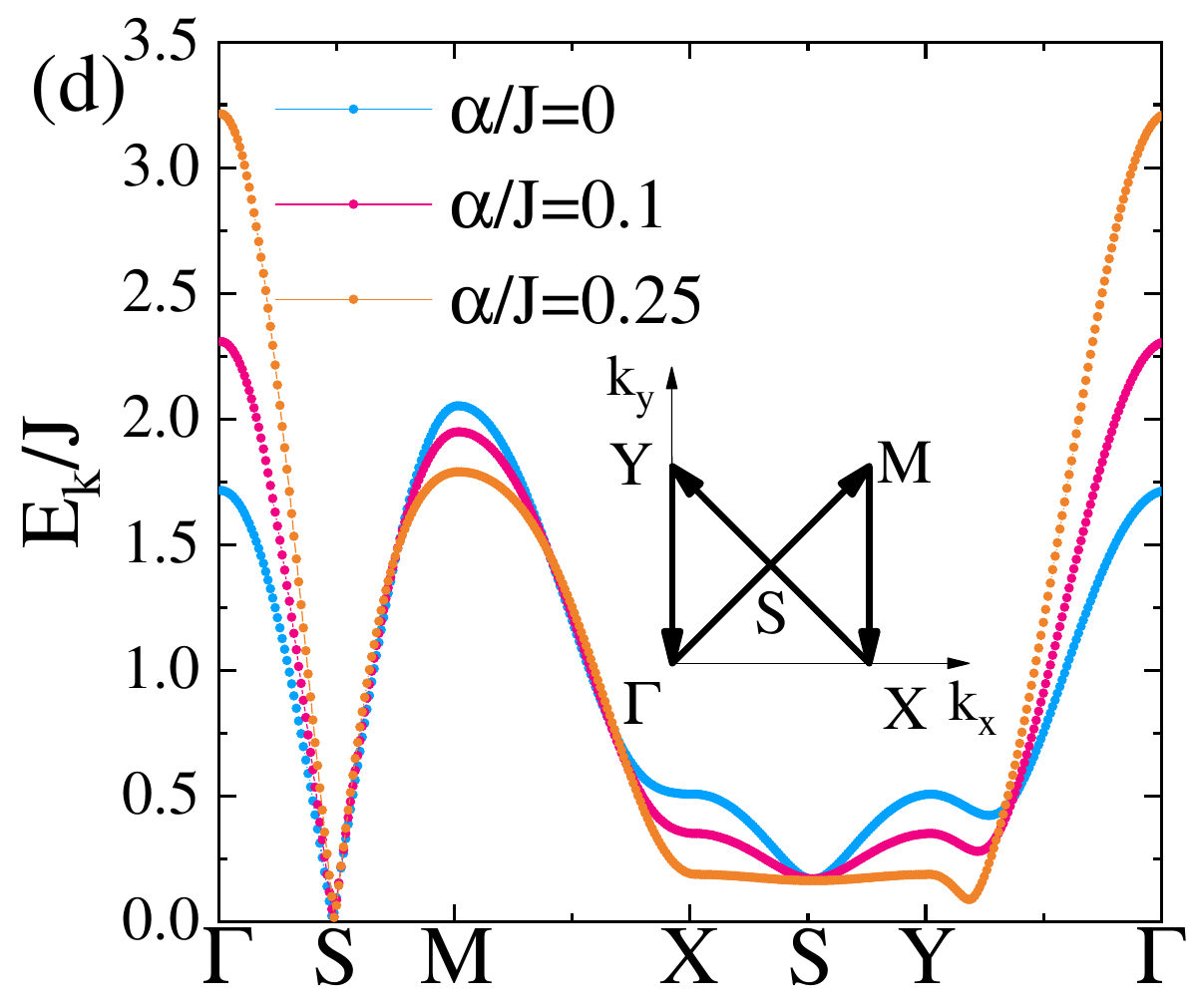}	
	\caption{\label{Figure3}(Color online)
	The quasiparticle dispersion $E_{\mathbf k}$ at doping concentration $\delta=0.125$ for typical values of $\alpha/J$. Here we consider $t/J=3$ and show the contour plot of $E_{\mathbf k}$ for (a) $\alpha/J=0$, (b) $\alpha/J=0.1$, (c) $\alpha/J=0.25$.  Panel (d) displays $E_{\mathbf{k}}$ along the high symmetry line, with the path illustrated in its inset.
	}
\end{figure*}

This paper is organized as follows. In Sec.~\ref{Sec2} we introduce the $t$-$J$-$ \alpha$ model and the SBMF approximation. In Sec.~\ref{Sec3}  we present the results obtained from the solution of self-consistent equations. Sec.~\ref{Sec4} is dedicated to showcasing the DMRG results of the $t$-$J$-$\alpha$  model, which are qualitatively in agreement with the SBMF findings on SC properties. Finally, in Sec.~\ref{Sec5} we discuss and in Sec.~\ref{Sec6} we summarize the main results.

\section{Model and Methods} \label{Sec2}
The one-band Hubbard model stands as the minimal model describing the correlated physics of doped Mott insulators. The Hamiltonian is given by
\begin{align}\label{Hubbard M}
	\mathcal{H}_{\mathrm{Hubbard}}=-t\sum_{\langle \mathbf{i},\mathbf{j} \rangle,\sigma}(c^{\dag}_{\mathbf{i}\sigma}c_{\mathbf{j}\sigma}+ \text{H.c.})+U\sum_{\mathbf i}n_{\mathbf{i}\uparrow}n_{\mathbf{i}\downarrow},
\end{align}
 where $c^{\dag}_{{\mathbf{i}}\sigma}$ and $c_{{\mathbf{i}}\sigma}$ are the electron operators of spin $\sigma=\uparrow,\downarrow$ and $n_{\mathbf{i}}=\sum_{\sigma}c^\dag_{\mathbf{i}\sigma}c_{\mathbf{i}\sigma}$ is the  density operator; $t$ is the hopping integral with summation  $\langle \mathbf{i},\mathbf{j}\rangle$ running over the nearest-neighbor (NN) sites; $U>0$ is the repulsive onsite Coulomb interaction. In the strong coupling limit ($U\gg t$), the {complete} effective Hamiltonian of the Hubbard model includes both the pure $t$-$J$ model and the three-site hopping term, as depicted in Fig.~\ref{Figure1}. The effective Hamiltonian~\cite{Auerbach1} reads
\begin{align}\label{Model}
\mathcal{H}_{\mathrm{\mathrm{t\small{-}J\small{-}\alpha}}}=\mathcal{P}(\mathcal{H}_{\mathrm{t}}+\mathcal{H}_{\mathrm{J}}+\mathcal{H}_{\mathrm{\alpha}})\mathcal{P},
\end{align}
where $\mathcal{P}$ is the projection operator onto the subspaces that eliminates doubly occupied sites.
The first term
 \begin{equation}
 \mathcal{H}_{\mathrm{t}}=-t\sum_{\langle \mathbf{i},\mathbf{j} \rangle,\sigma}(c^{\dag}_{\mathbf{i}\sigma}c_{\mathbf{j}\sigma}+ \text{H.c.})
 \end{equation}
 describes charge hopping with  amplitude $t$; the second term
 \begin{equation}
 \mathcal{H}_{\mathrm{J}}=J\sum_{\langle\mathbf{i},\mathbf{j}\rangle}(\mathbf{S}_{\mathbf{i}} \cdot \mathbf{S}_{\mathbf{j}} -\frac{1}{4} n_{\mathbf{i}} n_{\mathbf{j}})
 \end{equation}
 represents the superexchange interaction with exchange integral  $J={4t^2}/{U}$.  $\mathbf{S}_{\mathbf{i}}=\frac{1}{2}\sum_{\sigma\sigma'}c^{\dag}_{\mathbf{i}\sigma}\bm{\hat{\sigma}}_{\sigma\sigma'}c_{\mathbf{i}\sigma'} $ denotes the spin operator with Pauli matrix $\bm{\hat{\sigma}}$.
 The three-site hopping term $\mathcal{H}_{\mathrm{\alpha}}$, as sketched in Fig.~\ref{Figure1}(c), is written as
\begin{align}\label{H3s}
	\mathcal{H}_{\mathrm{\alpha}}=-\alpha\sum_{\substack{\langle \mathbf{i},\mathbf{j},\mathbf{k}\rangle,\sigma\\\mathbf{i}\ne \mathbf{k}}}(c^{\dag}_{\mathbf{i}\sigma}c^{\dag}_{\mathbf{j}\bar{\sigma}}c_{\mathbf{j}\bar{\sigma}}c_{\mathbf{k}\sigma}-c^{\dag}_{\mathbf{i}\sigma}c^{\dag}_{\mathbf{j}\bar{\sigma}}c_{\mathbf{j}\sigma}c_{\mathbf{k}\bar{\sigma}}).
\end{align}
Here, the three-site hopping amplitude $\alpha={t_{\mathbf{ij}}t_{\mathbf{jk}}}/{U}={J}/{4}$, where $t_{\mathbf{ij}}=t$ is the hopping integral for the NN sites $\mathbf{i}$ and $\mathbf{j}$,  is of the same order as superexchange interaction strength $J$. $\langle \mathbf{i},\mathbf{j}, \mathbf{k}\rangle$ denotes bonds of three-site hopping, with ${\mathbf i}\ne {\mathbf k}$ being neighbors of ${\mathbf j}$.  Here we set $J=1$ as the unit of energy, and tune $\alpha/J \in [0,0.25]$ to study the role of the three-site hopping term. Notably, $\alpha/J=0.25$ and $\alpha/J=0$ correspond to the effective model of large-$U$ Hubbard Hamiltonian and the pure $t$-$J$ model, respectively.

We use the SBMF method~\cite{SBMF1, SBMF2} to study the $t$-$J$-$ \alpha$ model on a square lattice written in Eq.~\eqref{Model} [see Appendix \ref{SecA2} for details].  Based on the resonating valence bond (RVB) theory~\cite{SingleH1, RVB1, RVB1.2}, the ground state can be approximated as a Gutzwiller projected BCS-like wave function. The electron annihilation operator $c_{\mathbf{i}\sigma}$ is decomposed as $c_{\mathbf{i}\sigma}= b^\dag_{\mathbf{i}}f_{\mathbf{i}\sigma}$, where $b^\dag_{\mathbf{i}}$ denotes bosonic holon creation operator and $f_{\mathbf{i}\sigma}$ represents fermionic spinon annihilation operator. Within the SBMF framework, the SC order is characterized by the condensation of holons and the RVB pairing of spinons. The local constraint $b_{\mathbf{i}}^{\dag}b_{\mathbf{i}}+\sum_{\sigma}f^{\dag}_{\mathbf{i}\sigma}f_{\mathbf{i}\sigma}=1$ is introduced with a static  Lagrange multiplier $\lambda_{\mathbf i}=\lambda$. Here we assume the simplest boson condensation  $\langle b_{\mathbf i}\rangle=\sqrt{\delta}$, where $\delta$ denotes doping concentration. The chemical potential
$\mu$  is determined by  $1-\delta=\langle \sum_{\sigma}f^{\dag}_{\mathbf{i}\sigma} f_{\mathbf{i}\sigma}\rangle$ under the local constraint. The density-density interaction $(1+b_{\mathbf i}^\dag b_{\mathbf i})(1+b_{\mathbf j}^\dag b_{\mathbf j})$ is approximated as $(1+\delta)^2\approx 1$. 
{In the framework of SBMF, it is natural to consider a doping-dependent amplitude for the three-site hopping term~\cite{Review2}, which has been adopted to argue the insignificant influence of this term at low-hole doping. However, it is pointed out that the effect of the three-site term becomes important at intermediate doping $\delta>0.1$~\cite{3s8}, which can be understood as the renormalization effect and is found to be beyond the order of $J\delta$~\cite{3s3,3s15}. In particular, starting from the three-band model, the strength of this term can be estimated as variable~\cite{Htre2, three-3s-3, three-band-new}, due to the detailed parameters including inter- and intra-orbital hopping amplitudes, orbital potential energies, and inter-orbital Coulomb interactions, etc. 
Therefore, we evaluate the role of this extended term by introducing an adjustable three-site term without doping dependence.  We also have examined the doping-dependent three-site hopping amplitude in SBMF solution, which gives qualitatively consistent conclusions but with a reduced impact. These results are presented in  Appendix~\ref{SecA1}. }
 
We give a Hartree-Fock-Bogoliubov factorization to decouple the Hamiltonian \eqref{Model} in the particle-hole  (i.e., hopping channel $f^{\dag}_{\mathbf{i}}f_{\mathbf{j}}$) and particle-particle  ( i.e., pairing channel $f^{\dag}_{\mathbf{i}}f^{\dag}_{\mathbf{j}}$)  channels~\cite{Review1}. 
We introduce the uniform bond order parameter $\chi_{\mathbf{ij}}=\langle  \sum_{\sigma}f^{\dag}_{\mathbf{i}\sigma}f_{\mathbf{j}\sigma}\rangle \equiv \chi$ and the RVB pairing order parameter $\Delta _{\mathbf{ij}}=\langle f_{\mathbf{i}\downarrow}f_{\mathbf{j}\uparrow}-f_{\mathbf{i}\uparrow}f_{\mathbf{j}\downarrow}\rangle\equiv\Delta_{x(y)}$ with $d$-wave pairing symmetry, i.e., $\Delta_{x}=-\Delta_{y} =  \Delta$. Here, $\Delta_{x(y)}$ represents NN bonds $\langle \mathbf{i},\mathbf{j}\rangle$ along $x(y)$-direction.
The resulting mean-field Hamiltonian can be obtained as
\begin{align}\label{HSBMF}
		\mathcal{H}_{\mathrm{SBMF}}=&\sum_{\mathbf{k},\sigma}\epsilon_{\mathbf{k}}  f^{\dag}_{\mathbf{k}\sigma}f_{\mathbf{k}\sigma}+\sum_{\mathbf{k}} \omega_{\mathbf{k}} b^\dag_{\mathbf{k}}b_{\mathbf{k}}\nonumber\\
		&-\sum_{\mathbf{k}} 	(\Delta^*_{\mathbf{k}} f_{\mathbf{k}\downarrow}f_{-\mathbf{k}\uparrow}+\Delta_{\mathbf{k}} f^\dag_{-\mathbf{k}\uparrow}f^\dag_{\mathbf{k}\downarrow}).
\end{align}
The corresponding free energy is given by
\begin{eqnarray}\label{Free}
	\mathcal{F}=&&\mathcal{H}_0+\sum_{\mathbf{k}} \epsilon_{\mathbf{k}}-\beta^{-1}\sum_{\mathbf{k}} 2\ln2\cosh\frac{\beta E_{\mathbf{k}}}{2}\nonumber\\
	&&+\beta^{-1}\sum_{\mathbf{k}} \ln(1-e^{-\beta \omega_{\mathbf{k}}}),
\end{eqnarray}
where $\mathcal{H}_0$ includes the constants  and  $\beta$ is the inverse of the temperature. The gap function is written as
\begin{align}\label{Delk}
	&&\Delta_{\mathbf{k}}=(J-2\alpha)\Delta\big(\cos k_{x}-\cos k_{y}\big).
\end{align}	
The spinon dispersion $\epsilon_{\mathbf{k}}$ and holon dispersion $\omega_{\mathbf{k}}$ read	
\begin{eqnarray}	
	\epsilon_{\mathbf{k}}&=&-[\mu_f +(\frac{1}{2}J\chi+3\alpha\chi+2t\delta)\big(\cos k_{x}+\cos k_{y} \big)\label{Epsk}\\
	&&+ \frac{1}{2}\alpha(1-\delta)\big(4\cos k_{x} \cos k_{y} +\cos2k_{x}+\cos2k_{y}\big)],\nonumber\\
	\omega_{\mathbf{k}}&=&\mu_b-2t\chi\big(\cos k_{x} +\cos k_{y}  \big).\label{Omek}
\end{eqnarray}		
Here $\mu_f$  and $\mu_b$  are chemical potentials for spinons and holons, respectively.
Both are determined self-consistently, as detailed in the self-consistent equations  Eq.~\eqref{nufse}, Eq.~\eqref{mufse} provided in Appendix~\ref{SecA2}. The mean-field order parameters can be obtained by searching for solutions that yield the lowest free energy. The corresponding self-consistent equations are given as follows
\begin{align}
	&&\Delta=\frac{1}{2N}\sum_{\mathbf{k}}\tanh\frac{\beta E_{\mathbf{k}}}{2}\frac{\Delta_{\mathbf{k}}}{E_{\mathbf{k}}}\big(\cos k_{x} -\cos k_{y} \big),\label{Deltasf}\\
	&&\chi=-\frac{1}{2N}\sum_{\mathbf{k}}\tanh\frac{\beta E_{\mathbf{k}}}{2}\frac{\epsilon_{\mathbf{k}}}{E_{\mathbf{k}}}\big(\cos k_{x} +\cos k_{y} \big).\label{Chisf}
\end{align}
Here $N$ is the total number of lattice sites. The quasiparticle energy can be obtained through $E_{\mathbf{k}}=\sqrt{\vert\Delta_{\mathbf{k}}\vert^2+\epsilon_{\mathbf{k}}^2}$.

\section{Mean Field Results} \label{Sec3}

\begin{figure}[b]%
	\includegraphics[width=0.48\textwidth]{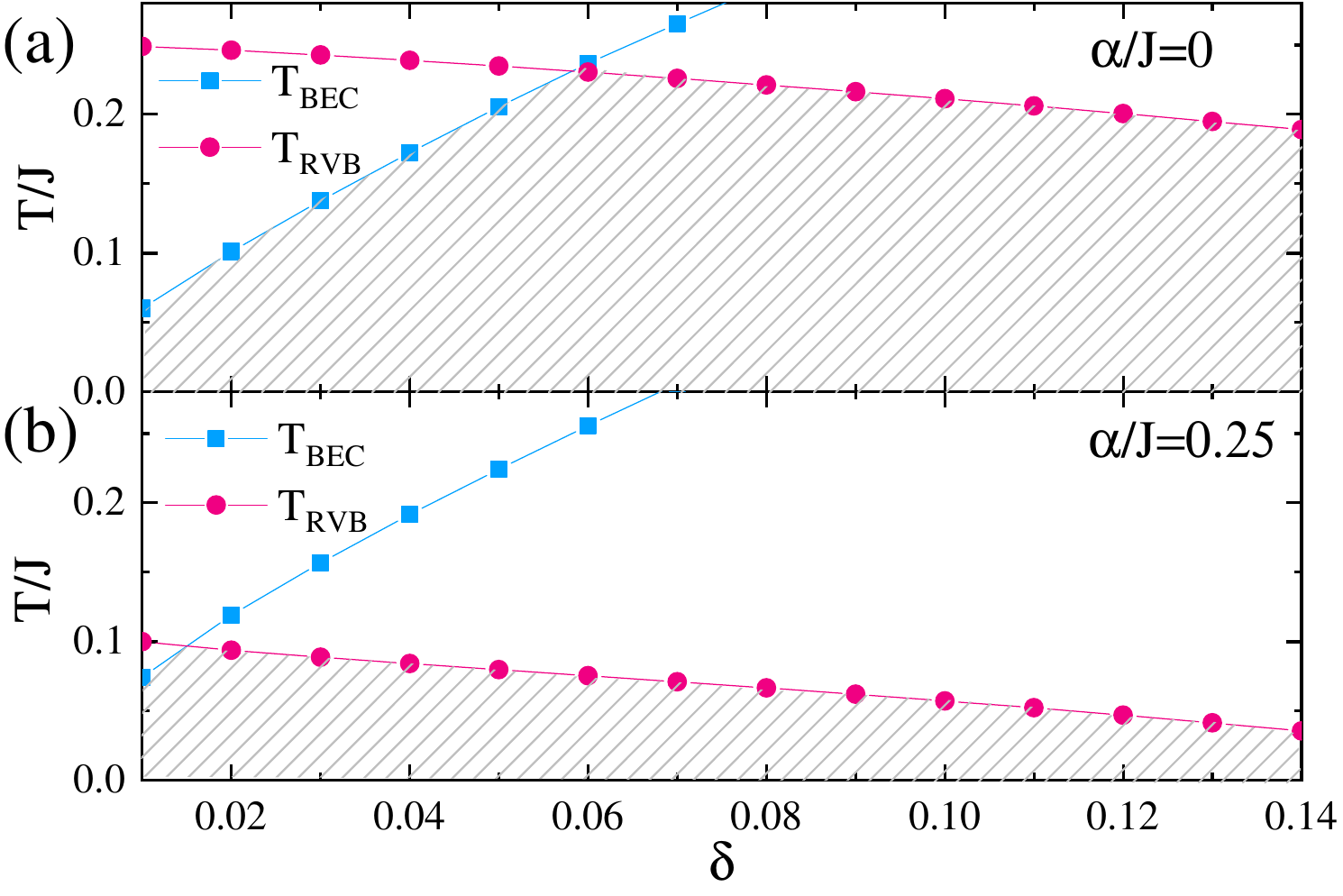}	 
	\caption{\label{Figure4}	(Color online)
The impact of the three-site hopping on the boson-condensation temperature $T_{\mathrm{BEC}}$ at different doping concentrations $\delta$.  Here we compare two cases: (a) the pure $t$-$J$ model with $\alpha/J=0$, and (b) the $t$-$J$-$\alpha$ model with $\alpha/J=0.25$, i.e., the effective Hamiltonian of the pure Hubbard model. The critical temperature $T_{\mathrm{BEC}}$ is calculated with $r_z=0.01$. The critical temperature $T_{\mathrm{RVB}}$ is also plotted to identify the phase diagram.  The shaded area represents the superconducting regime.
	}
\end{figure}

As schematically depicted in Fig.~\ref{Figure1}(c), the three-site hopping term $\mathcal{H}_{\mathrm{\alpha}}$ describes two types of hopping processes, with and without flipping the intermediate spin. The contributions from these processes are proportional to the doping concentration $\delta$. We can understand the three-site hopping as pair hopping of spin-singlet, which incorporates both collinear and non-collinear hopping shapes [see Fig.~\ref{Figure1}(b)].  The singlet hopping process can be expressed as follows~\cite{3s1}:
\begin{eqnarray} \label{Heff}
 \mathcal{H}_{\mathrm{\alpha}}=-\sum_{\substack{\langle \mathbf{i},\mathbf{j},\mathbf{k}\rangle\\\mathbf{i}\ne \mathbf{k}}}
 \alpha B^{\dag}_{\mathbf{i}\mathbf{j}}B_{\mathbf{k}\mathbf{j}},
\end{eqnarray}
where the spin-singlet pairing operator is defined as $B^{\dag}_{\mathbf{i}\mathbf{j}}=c^{\dag}_{\mathbf{i}\uparrow}c^{\dag}_{\mathbf{j}\downarrow}-c^{\dag}_{\mathbf{i}\downarrow}c^{\dag}_{\mathbf{j}\uparrow}$.
By implementing a Hartree-Fock factorization on Eq.~\eqref{Heff}, it is straightforward to notice that the nonzero $\langle B^\dag_{\mathbf{i}\mathbf{j}}\rangle$ arising from singlet hopping impacts the mean-field parameter $\Delta$, which is dominated by the superexchange mechanism in a pure $t$-$J$ model~\cite{SBMF1, SBMF2}.
At the mean-field level, the phase diagram of $t$-$J$-$\alpha$ should be influenced by both the superexchange interaction and pair singlet hopping.

We first discuss the solutions for $d$-wave RVB pairing as a function of doping concentration $\delta$. The  RVB pairing order parameter $\Delta$ is displayed in Fig.~\ref{Figure2}(a) at $T=0$ and the corresponding critical temperature  $T_{\mathrm{RVB}}$~\cite{SBMF1, SBMF2} is plotted in Fig.~\ref{Figure2}(b).  {The SBMF solution with a doping-dependent three-site hopping amplitude is provided in Appendix~\ref{SecA1}}. The $d$-wave RVB pairing is suppressed as the three-site hopping amplitude $\alpha$ increases.
We further investigate the solutions for $d$-wave RVB pairing as a function of ratio $t/J$. At a fixed doping concentration $\delta=0.1$, as shown in Fig.~\ref{Figure2}(c), $T_{\mathrm{RVB}}$ is suppressed by $\alpha$ and $t/J$ but it is still nonzero over a broad range of coupling strength $t/J$$\approx$ $2\sim6$.  The mean-field solutions demonstrate the suppression of SC order by incorporating the three-site hopping term. Due to the presence of singlet pair hopping processes, the coefficient of the gap function $\Delta_{\mathbf{k}}$ is $(J-2\alpha)$[see Eq.~\eqref{Delk}]. Since $\alpha>0$, $\mathcal{H}_{\mathrm{\alpha}}$ competes with  $\mathcal{H}_{\mathrm{J}}$, leading to the general suppression of the superexchange mechanism. This provides an alternative understanding of the numerical discrepancies between the weak or absent superconductivity in the pure Hubbard model and the strong superconductivity in the pure $t$-$J$ model.

%%%%%%%%%%%%%%%%%%%%%%%%%%%%%%%%%%%%%%%%%%%%%%%%%%%
\begin{figure}[tbp]
\begin{center}
\includegraphics[width=0.48\textwidth]{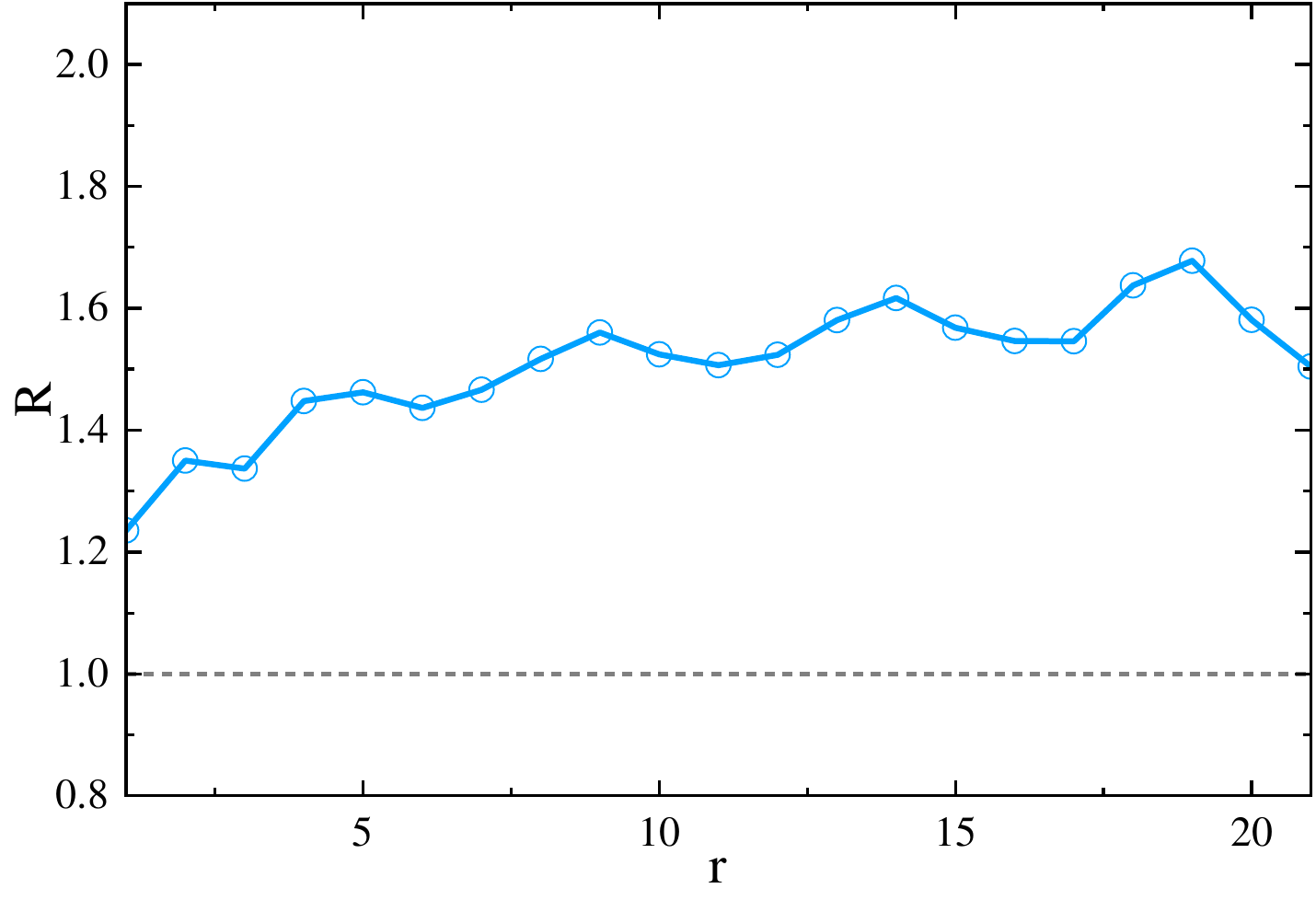}
\end{center}
\par
\caption{\label{Figure5}(Color online)
Ratio $R\equiv D_{yy}(r)|_{\alpha/J=0}/D_{yy}(r)|_{\alpha/J=0.25}$ from DMRG simulation. Here, we set $\delta=0.1$, $t/J=3$, $N=30\times 4$. 
}
\end{figure}
%%%%%%%%%%%%%%%%%%%%%%%%%%%%%%%%%%%%%%%%%%%%%%%%%%%

We further examine the effect of three-site hopping on quasiparticle dispersion $E_{\mathbf{k}}$.   The contour plots are presented in Figs.~\ref{Figure3}(a-c) for $\alpha/J=0$, $0.1$, $0.25$ at doping concentration $\delta=0.125$. And in Fig.~\ref{Figure3}(d), the corresponding dispersion is plotted along the high symmetry line [illustrated in the inset of  Fig.~\ref{Figure3} (d)]. The quasiparticle dispersion illustrated in Fig.~\ref{Figure3}(d) has a local minimum around $S$ point (i.e., ${\mathbf {k}}=(\frac{\pi}{2},\frac{\pi}{2})$), corresponding to the valley near $S$ point in Figs.~\ref{Figure3}(a-c), and displays nearly flat quasiparticle energy along the $X$-to-$Y$ line.
Along the $X$-to-$Y$ line, increasing $\alpha$ further flattens the dispersion [see Fig.~\ref{Figure3}(d)], as suggested by the energy difference $\vert E_{{\mathbf k}=X}-E_{{\mathbf k}=S}\vert$, and enlarges the range of the valley [see Figs.~\ref{Figure3}(a-c)]. Due to the broadening of the valley along the $X$-to-$Y$ line, the dispersion along the $Y$-to-$\Gamma$ line [plotted in Fig.~\ref{Figure3}(d)] shows that a local minimum approaches to $Y$ as $\alpha$ increases.  The results in Fig.~\ref{Figure3} indicate that increasing $\alpha$ diminishes the feature of the valley near $S$, and when $\Delta$ approaches zero, the valley reduces to contours nearly parallel to the $X$-to-$Y$ line. In addition, as presented in Fig.~\ref{Figure3}, the quasiparticle energy increases at $\Gamma$ point (i.e., ${\mathbf {k}}=(0,0)$)  with growing $\alpha$ and decreases at  $X$ point (i.e., ${\mathbf {k}}=(\pi,0)$) and $Y$ point (i.e., ${\mathbf {k}}=(0,\pi)$). Due to the formulas of $\Delta_k$ and $\epsilon_k$, the increase corresponds to the rising spinon energy at $\Gamma$, and the decrease is attributed to the suppression of the $d$-wave pairing. When $\Delta$ is suppressed to nearly zero by increasing $\alpha$, the spinon dispersion $\epsilon_{\mathbf k}$  dominates the quasiparticle energy. Mean-field analysis indicates that incorporating the three-site hopping term modifies the shape of quasiparticle dispersion $E_{\mathbf k}$.  As a result of suppressed $d$-wave pairing, the feature of the valley near the $S$ point is diminished, and the spinon energy becomes more significant in quasiparticle energy. {The quasiparticle dispersion $E_{\mathbf{k}}$ using the doping-dependent three-site hopping amplitude is presented in Appendix~\ref{SecA1}}.

\begin{figure*}[tbp]%

		\includegraphics[width=0.48\textwidth]{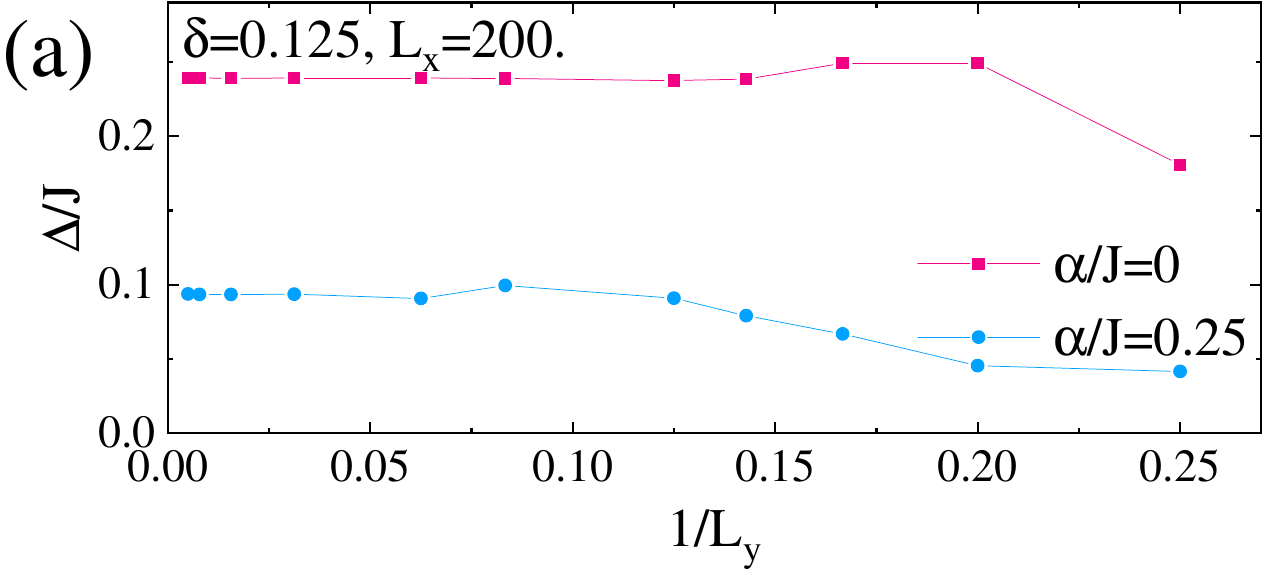}	
		\includegraphics[width=0.48\textwidth]{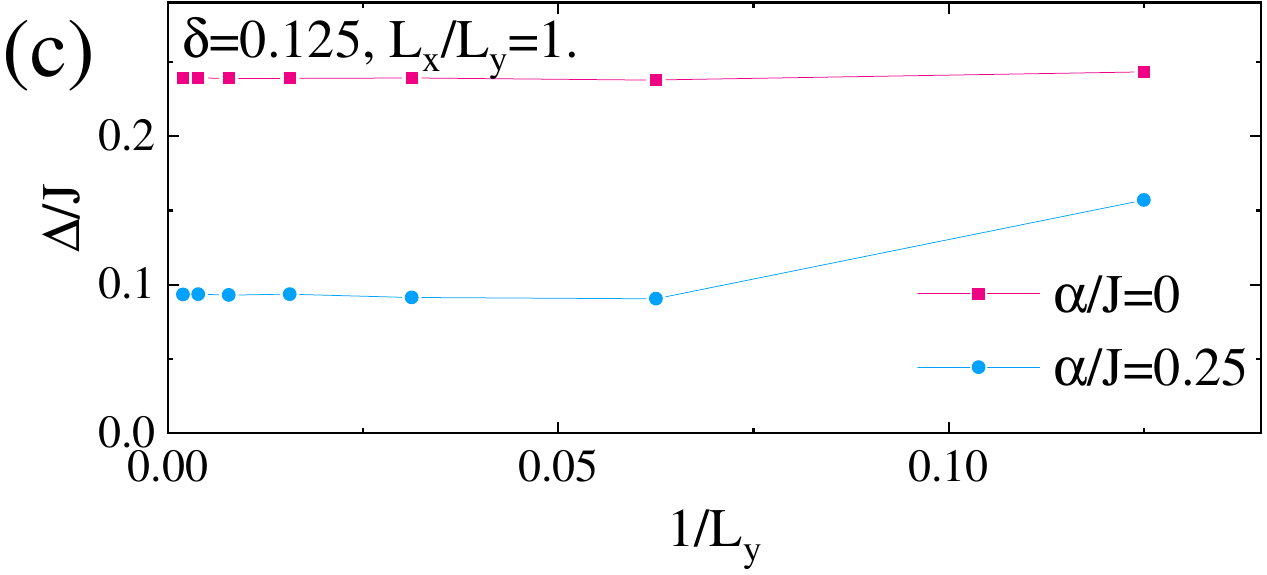}	
  		\includegraphics[width=0.48\textwidth]{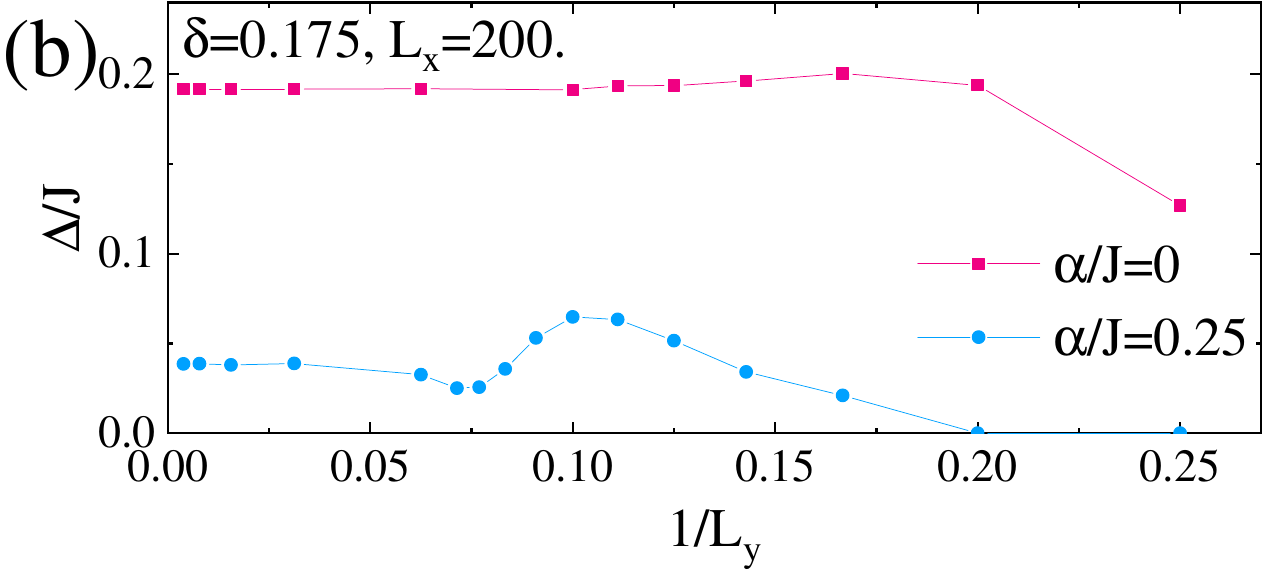}	
    		\includegraphics[width=0.48\textwidth]{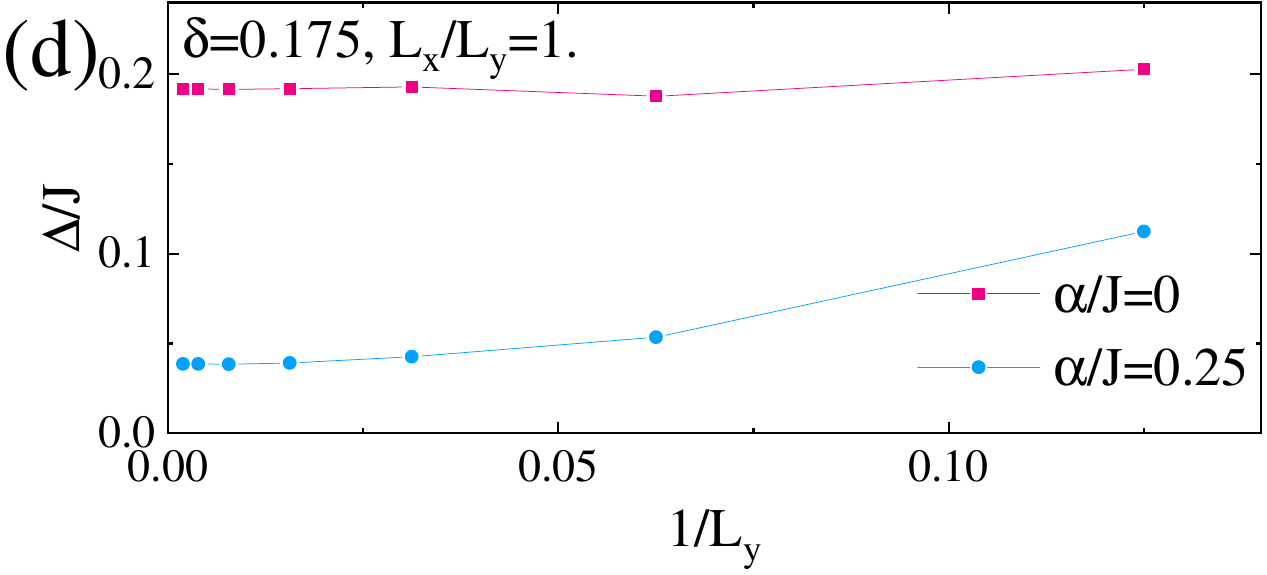}	

	\par

	\caption{\label{Figure6}(Color online)
		 {The superconducting order parameter $\Delta$ versus the inverse of width $L_y$ in the $t$-$J$-$\alpha$ model. Here, we consider two cases:  the pure $t$-$J$ model with $\alpha/J=0$ (squares) and the effective model of the pure Hubbard Hamiltonian with $\alpha/J=0.25$ (circles). The doping concentrations are set at $\delta=0.125$ (a,c) and $\delta=0.175$ (b,d) with a fixed ratio $t/J=3$. System sizes are $N=L_x\times L_y$, where $L_x$ and $L_y$ denote the system length and width, respectively. The aspect ratio is $r\equiv L_x/L_y$. In panels (a-b), we fix $L_x=200$  and vary aspect ratio $r$ down to $1$.  In panels (c-d), we fix $r=1$ and vary system size $N$ up to $200\times200$.}
	}
\end{figure*}

Next, we investigate the influence of three-site hopping on the boson-condensation temperature $T_{\mathrm{BEC}}$, which also determines the phase diagram~\cite{Review1,Review2}.  The SC state is characterized by both $\langle b_{\mathbf{i}}b_{\mathbf{j}} \rangle\ne0$ and $\langle f_{\mathbf{i}\sigma}f_{\mathbf{j}\bar{\sigma}} \rangle\ne0$, corresponding respectively to the critical temperatures $T_{\mathrm{BEC}}$ and $T_{\mathrm{RVB}}$. The critical temperature of superconductivity $T_{\mathrm{c}}$ is determined from $T_{\mathrm{c}}=\min\{T_{\mathrm{RVB}}, T_{\mathrm{BEC}}\}$. The nonvanishing  order parameter $\langle b_{\mathbf{i}}b_{\mathbf{j}} \rangle$ is approximately obtained by single boson condensation $\langle b_{\mathbf{i}}\rangle =\sqrt{\delta}$. In order to get a finite critical temperature $T_{\mathrm{BEC}}$, it is necessary to introduce a weak NN three-dimensional hopping between layers~\cite{SBMF1, SBMF2,SBMF3}. The self-consistent equations are solved by assuming a small interlayer hopping amplitude $t_{z}$ in the $z$ direction to obtain a finite boson condensation temperature $T_{\mathrm{BEC}}$ and $T_{\mathrm{RVB}}$. We set ratio $r_{z}=t_{z}/t$, assuming exchange coupling along the $z$-direction as $J_z=r_z^2J$. Singlet-pair hopping from $xy$-plane to the $z$-direction is denoted as $\alpha_{z-xy}=r_z\alpha$, and along the $z$-direction as $\alpha_{z-z}=r_z^2\alpha$.  We can then update the gap function and the energy dispersion to the following expressions:
	\begin{eqnarray}	
		&&\Delta_{\mathbf{k}}=(J-2\alpha)\Delta\big(\cos k_{x}-\cos k_{y}\big)\\
		&&+4\alpha r_z\Delta_z\big(\cos k_{x}+\cos k_{y}\big)+(J+2\alpha)(r_z)^2\Delta_z\cos k_{z},\nonumber \\
	&&\epsilon_{\mathbf{k}}=-[\mu_f+(\frac{1}{2}J\chi-\alpha\chi)\big(\cos k_{x}+\cos k_{y}+(r_z)^2\cos k_{z} \big)\nonumber \\
&& +2(2+r_z)\alpha\chi\big(\cos k_{x}+\cos k_{y}+r_z\cos k_{z} \big)\nonumber\\
&&+2t\delta\big(\cos k_{x}+\cos k_{y}+r_z\cos k_{z} \big)\nonumber\\
&&+ 2\alpha(1-\delta)\big(\cos k_{x} \cos k_{y}+r_z[\cos k_{y} \cos k_{z}+\cos k_{z} \cos k_{x}]\big)\nonumber\\
&&+\frac{1}{2}\alpha(1-\delta)\big( \cos2k_{x}+\cos2k_{y}+(r_z)^2\cos2k_{z}\big)],\\
		&&\omega_{\mathbf{k}}=\mu_b-2t\chi\big(\cos k_{x} +\cos k_{y}+r_z\cos k_{z}  \big),
	\end{eqnarray}	
	where $\Delta_z= \Delta$ is the NN bonds along $z$-direction. The self-consistent equations are solved with $\mu_b=0$~\cite{SBMF2}. In Fig.~\ref{Figure4} we show the influence of $\alpha$ on the critical temperature  $T_{\mathrm{BEC}}$ with $r_{z}=0.01$, and  the shaded area represents the superconducting regime.  {The influence of a doping-dependent three-site hopping amplitude is also estimated in Appendix~\ref{SecA1}}.  Comparing results for $\alpha/J=0$ and $0.25$, the critical temperature $T_{\mathrm{BEC}}$ is slightly enhanced by the presence of $\mathcal{H}_{\mathrm{\alpha}}$, in sharp contrast to the strong suppression of $T_{\mathrm{RVB}}$. The mean-field analysis indicates that the impact of this term on the phase diagram is primarily mediated by altering RVB pairing.

\section{NUMERICAL Results}\label{Sec4}

In this section, we employ DMRG~\cite{NDMRG1, NDMRG2, NDMRG3} to confirm the validity of SBMF results.
DMRG has been proven to be one of the most accurate methods to obtain the ground-state properties of many-body systems, but its computational cost grows exponentially with system width, therefore we focus on the quasi-one-dimensional cylinders as usual practice~\cite{NDMRG4}
We compute Cooper pair correlations to examine the properties of superconductivity.
In the quasi-one dimension, true long-range order in the pair correlation function is forbidden by the Mermin-Wagner theorem, and then the pair correlations decay in a power-law fashion. We employ a square lattice geometry, defined by the primitive vectors ${\boldsymbol{e}_{{x}}}=(1,0)$, $\boldsymbol{e}_{{y}}=(0,1)$, and wrapped on cylinders with a lattice spacing of unity. The system size is denoted as $N = L_x \times L_y$, where $L_x$ and $L_y$ correspond to the cylinder length and circumference, respectively. In our study, we focus on the width-4 cylinders, i.e., $L_y=4$. The pair correlations are defined as
\begin{equation}\label{eq:D}
D_{\alpha\beta}(\mathbf{r}) \equiv \left\langle \hat{\Delta}^{\dagger}_{\alpha}(\mathbf{i}_0) \hat{\Delta}_{\beta}(\mathbf{i}_0 +\mathbf{r})\right\rangle,
\end{equation}
where the pair operator is given by $\hat{\Delta}_{\alpha}(\mathbf{i}) \equiv \frac 1 {\sqrt{2}}\sum_{\sigma}\sigma {c}_ {\mathbf{i}, \sigma} {c}_ {\mathbf{i}+\boldsymbol{e}_\alpha, \bar{\sigma}}$, and $\alpha, \beta = x, y$. When calculating the correlations, we set the reference position $\mathbf{i}_0=(L_x/4,y_0)$ with $y_0=2$ to avoid a boundary effect. In quasi-one-dimensional cylinders, we explore the presence of quasi-long-range order, which is characterized by $D_{\alpha\beta}({r})\sim r^{-\eta_{\mathrm{sc}}}$. Here, we set $\mathbf{r}=r\boldsymbol{e}_x$. Specifically, $\eta_{\mathrm{sc}} < 2$ indicates a divergent superconducting susceptibility in two dimensions as the temperature $T\to 0$.

Figure~\ref{Figure5} illustrates the ratio $R$ of $D_{yy}(r)$ at $\alpha/J=0$ compared to $\alpha/J=0.25$, i.e., $R\equiv D_{yy}(r)|_{\alpha/J=0}/D_{yy}(r)|_{\alpha/J=0.25}$. {We find that $R$ consistently exceeds 1, suggesting that pair correlation is relatively suppressed by including the three-site hopping term $H_\alpha$. We also notice that this suppressive effect is not prominent in DMRG cylinders. } Moreover, we find exponentially decaying spin correlations $S(r)\equiv \langle \mathbf{S}_{{\mathbf{i}}_0}\cdot \mathbf{S}_{{\mathbf{i}_0+r\boldsymbol{e}_x}}\rangle$ and single-particle propagator $C(r)\equiv\sum_{\sigma}\langle c_{\mathbf{i}_0,\sigma}^\dagger c_{{\mathbf{i}_0+r\boldsymbol{e}_x},\sigma}\rangle $ [see Appendix~\ref{SecA4}]. This underscores the persistence of qualitative decay behaviors across various correlations in the presence of the three-site hopping term.

\section{Discussion} \label{Sec5}

\begin{figure}[tbp]%
	\includegraphics[width=0.48\textwidth]{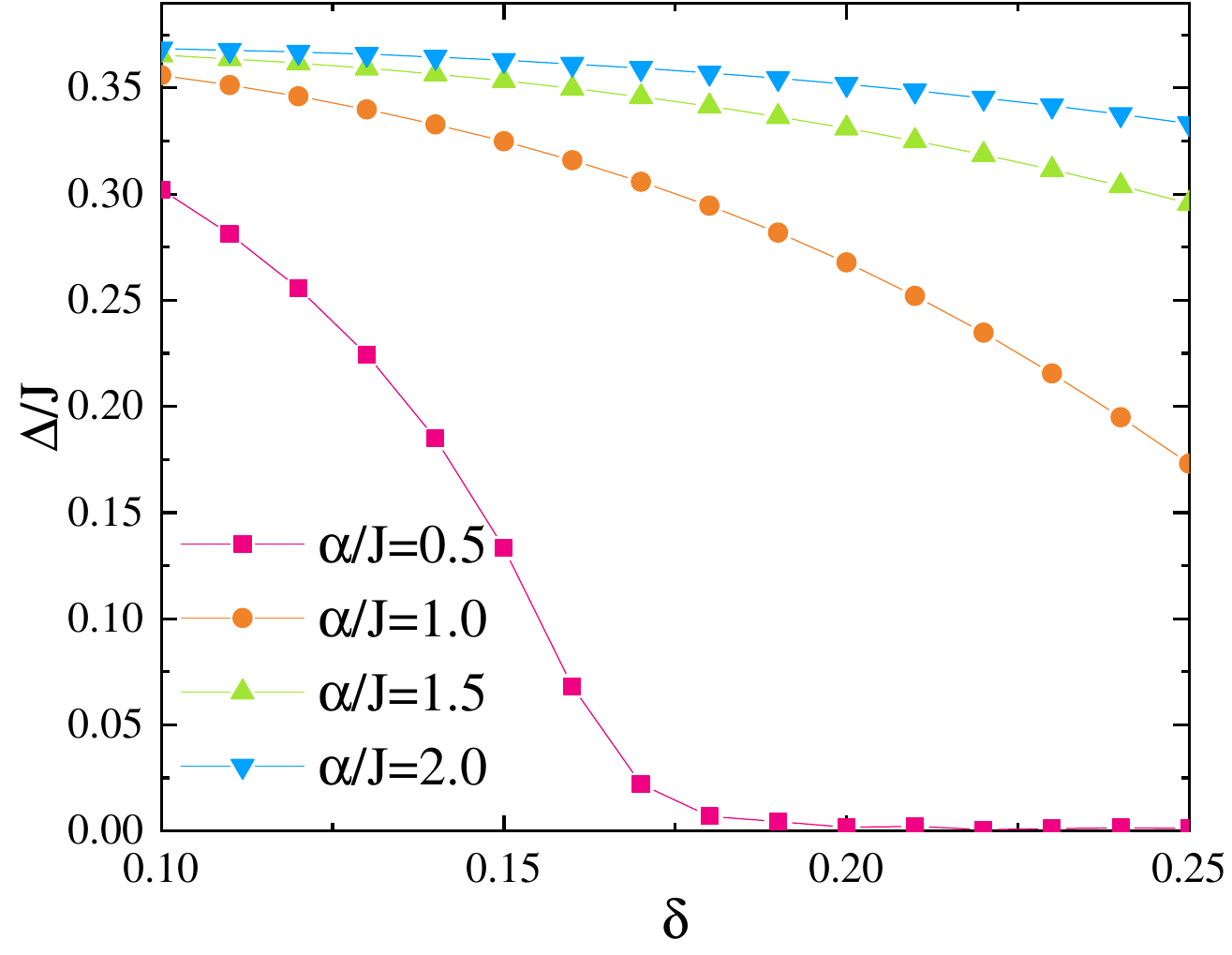}				
	\caption{\label{Figure7}(color online)
		For larger three-site hopping amplitude $\alpha$, the $s$-wave pairing order parameter $\Delta$ as a function of doping concentration $\delta$.  Here, we fix $t/J=3$ and consider $\alpha/J=0.5,1,1.5,2$, where the $d$-wave RVB pairing vanishes. 	
	}
\end{figure}

Inspired by recent debates in DMRG studies regarding the existence of superconductivity in the pure Hubbard and pure $t$-$J$ models, and considering the limitations posed by conducting DMRG calculations on lattices with finite widths---a factor previously emphasized for its crucial role in detecting SC order---we examine the impact of system size and aspect ratio of the square lattice on SC order. Specifically, we conduct the SBMF analysis on square lattices with system sizes accessible to DMRG calculations, and compare the superconducting order parameter $\Delta$ between the pure $t$-$J$ model and the effective model of the pure Hubbard Hamiltonian, which corresponds to the $t$-$J$-$\alpha$ model with $\alpha=0$ and $\alpha=0.25$, respectively. {The SBMF solution with a doping-dependent three-site hopping amplitude is provided in Appendix~\ref{SecA1}}. Here, we set the aspect ratio as $r\equiv L_x/L_y$, with $L_x$ and $L_y$ representing the system length and width. Generally, $r\gg1$ for DMRG calculations.

In Fig.~\ref{Figure6}(a-b), we fix $L_x=200$ and vary aspect ratio $r$ from $r\gg 1$ down to $r=1$. The convergence of $\Delta$ with increasing $L_y$ for $\alpha/J=0.25$ is much slower than that for $\alpha/J=0$. For $\alpha/J=0.25$, $\Delta$ shows stronger finite size effect when {$r\gg 1$}, and the saturated value of  $\Delta$ decreases with increasing doping concentration $\delta$. In particular, the superconductivity almost vanishes at $\delta=0.175$ for a larger aspect ratio. In Fig.~\ref{Figure6}(c-d), we fix $r=1$ and vary system size $N$ up to $200\times200$.
The SC order parameter $\Delta$ is also larger at $\alpha/J=0$ than $\alpha/J=0.25$. $\Delta$ saturates faster with increasing system size for $\alpha/J=0$ and its amplitude decreases with increasing doping concentration $\delta$. The mean-field findings indicate the effective model of the pure Hubbard Hamiltonian is more sensitive to system size and aspect ratio. Considering the order parameter is also suppressed compared with the pure $t$-$J$ model, this observation suggests a greater challenge to probe superconductivity numerically within the effective model of the pure Hubbard Hamiltonian.

Within the framework of mean field theory, one should also consider the possibility of $s$-wave-like solution  [see details in Appendix~\ref{SecA3}]~\cite{ap201,3s8,3s11}{, i.e., $\Delta_{x}=\Delta_{y} =  \Delta$}.  In a pure $t$-$J$ model, the $d$-wave pairing state is always favored. However, the three-site hopping term favors the extended $s$-wave component~\cite{ap201,3s8,3s11}.
We tune the three-site hopping amplitude $\alpha$ beyond the physical region to explore the possible $s$-wave solution, and the results are presented in Fig.~\ref{Figure7}.  {The results of large $\alpha$ for a doping-dependent three-site hopping amplitude are provided in Appendix~\ref{SecA1}}.  The $d$-wave order parameters are strongly suppressed to zero at $\alpha/J=0.5,1.0,1.5$, and $2.0$. In contrast, the $s$-wave pairing order parameter is enhanced with increasing $\alpha$.
{We tune the three-site hopping amplitude $\alpha$ beyond the physical region to explore the possible $s$-wave solution, and find that the $s$-wave pairing order parameter is enhanced with increasing $\alpha$, as illustrated in Fig.~\ref{Figure7}. By contrast, the $d$-wave order parameters are strongly suppressed to zero at $\alpha/J=0.5$, $ 1.0$, $1.5$.}

When comparing the model study with real materials like cuprates, it is also proposed to start from the three-band Hubbard model and its effective Hamiltonian in the strong coupling limit.~\cite{three-band-new} In this case, the exact form of the three-site hopping term derived from the three-band model is similar to the single-band model~\cite{Htre2}. In the strong coupling limit, where the energy potential between the oxygen site and the copper site is large, one can obtain a more general three-site term~\cite{Htre2},
	\begin{align}
	\mathcal{H}_{\mathrm{\alpha}}=-\sum_{\substack{\langle \mathbf{i},\mathbf{j},\mathbf{k}\rangle,\sigma\\\mathbf{i}\ne \mathbf{k}}}(\alpha_1c^{\dag}_{\mathbf{i}\sigma}c^{\dag}_{\mathbf{j}\bar{\sigma}}c_{\mathbf{j}\bar{\sigma}}c_{\mathbf{k}\sigma}-\alpha_2c^{\dag}_{\mathbf{i}\sigma}c^{\dag}_{\mathbf{j}\bar{\sigma}}c_{\mathbf{j}\sigma}c_{\mathbf{k}\bar{\sigma}}),
\end{align}
where the strengths of the two hopping processes, as depicted in Fig.~\ref{Figure1}, are considered different. They are determined by the detailed potentials and interactions of the three-band model. Under mean-field approximations, the contributions of the three-site term into the pairing channel are qualitatively consistent with the case derived from the one-band model. The corresponding Hamiltonian is formulated as follows:
	\begin{eqnarray}
		\mathcal{H}^{\Delta}_{\mathrm{\alpha}}=&&-\sum_{\substack{\langle \mathbf{i},\mathbf{j},\mathbf{k}\rangle\\\mathbf{i}\ne \mathbf{k}}}\alpha[\Delta^*_{\mathbf{i}\mathbf{j}}(f_{\mathbf{j}\downarrow}f_{\mathbf{k}\uparrow}-f_{\mathbf{j}\uparrow}f_{\mathbf{k}\downarrow})\nonumber\\
  &&+\Delta_{\mathbf{j}\mathbf{k}}(f^\dag_{\mathbf{i}\uparrow}f^\dag_{\mathbf{j}\downarrow}-f^\dag_{\mathbf{i}\downarrow}f^\dag_{\mathbf{j}\uparrow})],
	\end{eqnarray}	
	where we set $\alpha=(\alpha_1+\alpha_2)/{2}$. When the detailed potentials and interactions of the three-band model are varied, $\alpha$ can be considered adjustable. Thus, employing a tunable $\alpha$ to assess the effects of the three-site hopping provides a practical and reasonable approximation. As we focus on a large value of $t/J$ and we focus on the superconducting properties which are mainly determined by the pairing channel, the one-band model with tunable coefficients in the three-site hopping term potentially captures the essential physics of the cuprates.

 The NN Coulomb interaction is known to play an important role in the superconductivity of cuprates~\cite{Ht5}. At the mean-field level, the related gap function can be expressed as
\begin{eqnarray}
	\Delta_{\mathbf{k}}=	&&4\alpha (\Delta_{x}+\Delta_{y})(\cos(k_x)+\cos(k_y))\nonumber\\
 &&+2(J_{\Delta}-\alpha)(\Delta_x\cos(k_x)+\Delta_y\cos(k_y)).
\end{eqnarray}
where $J_{\Delta}=\frac{1}{2}(J-V)$,  $J$ is the strength for superexchange term $J\sum_{\langle \mathbf{i} \mathbf{j}\rangle}(\mathbf{S}_{\mathbf{i}}\cdot\mathbf{S}_{\mathbf{j}}-\frac{1}{4}n_{\mathbf{i}}n_{\mathbf{j}})$, and  $V$ is the strength for NN Coulomb interaction term $V\sum_{\langle \mathbf{i} \mathbf{j}\rangle}n_{\mathbf{i}}n_{\mathbf{j}}$. We can find that the repulsive NN Coulomb interaction ($V>0$) disfavors SC pairing at the mean-field level. The recent study suggests that including fluctuations beyond the mean-field level screens the NN Coulomb interaction, thereby preserving $d$-wave superconductivity~\cite{exmfn}.
In our work, both the mean-field analysis and DMRG numerical calculations suggest the weakening of the $d$-wave superconductivity by the three-site hopping term. The impact of there-site hopping on SC order is different from the impact of the fluctuations combined with Coulomb repulsion reported in Ref.~~\cite{exmfn}. A systematical understanding of the interplay between these two effects could be intriguing and calls for systematic study in the future.

\section{Conclusion} \label{Sec6}
In this work, we combine slave-boson mean-field (SBMF) theory and density matrix renormalization group (DMRG) to study the $t$-$J$-$\alpha$ model and provide a comprehensive analysis of the role of the three-site hopping term $\mathcal{H}_{\mathrm{\alpha}}$ in superconductivity emerged in doped Mott insulators on the square lattice. Our study employs the strengths of the slave-boson mean-field theory in microscopically understanding the impact of the three-site hopping term on the superconducting order parameter, and the capacity of DMRG in capturing the superconductivity. We also vary the three-site hopping amplitude $\alpha$, discuss the  $s$-wave solution at high $\alpha$ values, assess the impact of the three-site term on quasiparticle dispersion and the boson-condensation temperature and explore the impact of systems sizes and aspect ratios on superconductivity. The mean field analysis suggests the suppression of $d$-wave superconductivity in the presence of a three-site hopping term, consistent with numerical observations by DMRG. This suppression could be understood as a result of competition between superexchange interaction and three-site hopping (i.e., singlet pair hopping), the former favors $d$-wave pairing while the latter favors $s$-wave pairing.
Our findings may offer an alternative understanding of the recent numerical contrasting findings in the strong coupling regime: the absent or weak superconductivity in the pure Hubbard model, versus the robust superconductivity in the $t$-$J$ model without including the three-site hopping term.

In particular, our comparative study of the $t$-$J$ and $t$-$J$-$\alpha$ model on other physical phenomena may also provide insights into the difference between the pure Hubbard and pure $t$-$J$ models, and stimulate future studies in identifying the role of three-site hopping in doped Mott insulators besides the suppression of $d$-wave superconductivity. Moreover, our work may also stimulate future studies in identifying the role of the three-site hopping term in superconductivity on other lattice geometries, such as the triangular lattice~\cite{Zhu2022Doped,Chen2022Proposal, KevinHuang2022, Huang2023Quantum, ZhengZhu2023}, where superconductivity is recently discovered in materials like twisted bilayer of transition-metal-dichalcogenides~\cite{xia2024unconventional, guo2024superconductivity}.

\appendix

\begin{acknowledgments}
We thank the helpful discussions with Fuchun Zhang, Shuai Chen.
This work is supported by National Natural Science Foundation of China (Grant No.12074375), the Fundamental Research Funds for the Central Universities and the Strategic Priority Research Program of CAS (Grant No.XDB33000000).
\end{acknowledgments}

\begin{figure}[tbp]%
	\includegraphics[width=0.48\textwidth]{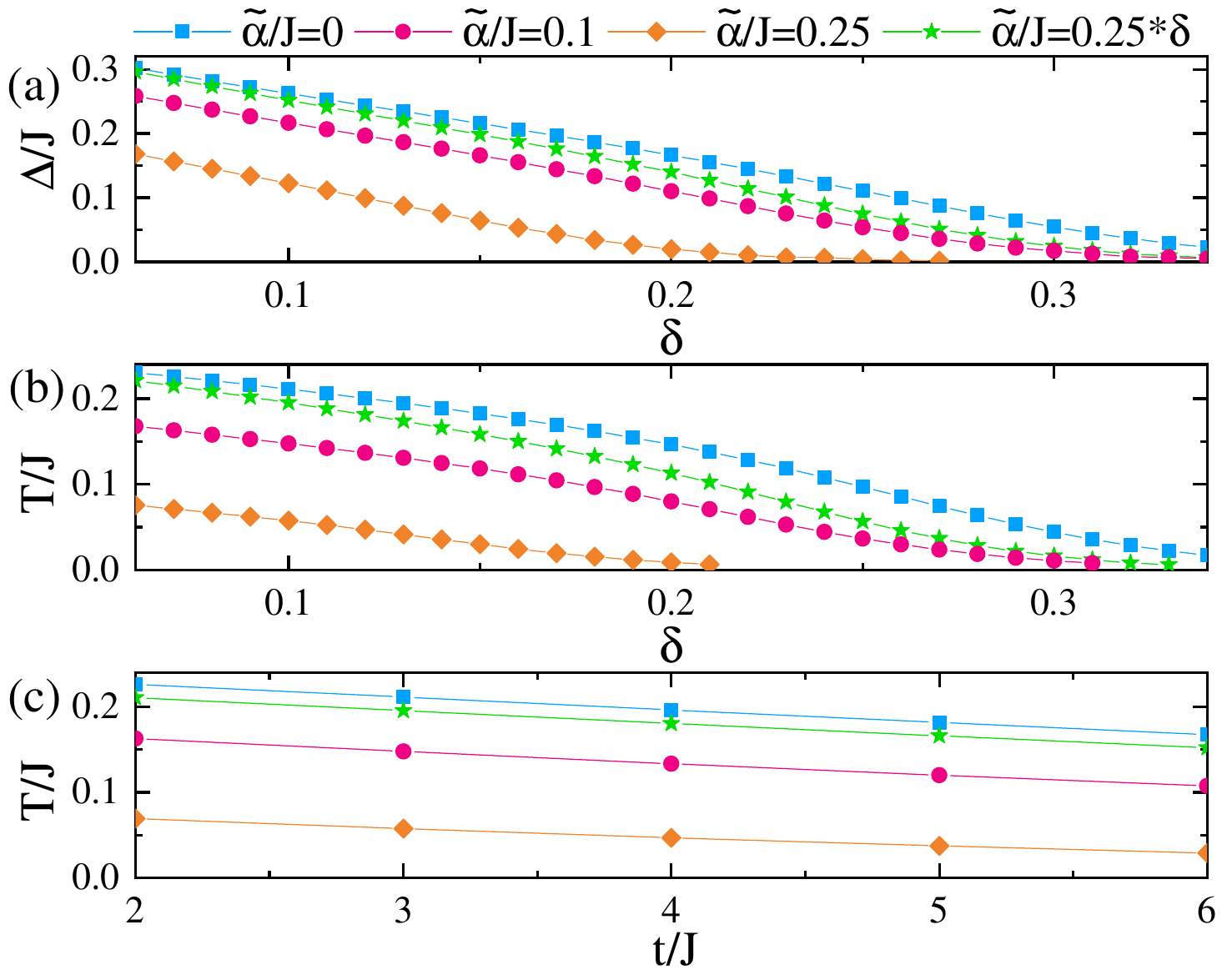}		
	\caption{\label{Figure8}(Color online)
{
Suppression of the $d$-wave superconductivity with the increase of the effective three-site hopping amplitude $\tilde{\alpha}$. Here, we consider $t/J=3$ and compare the results of the doping-dependent effective amplitude $\tilde{\alpha}/J=0.25\delta$ with those presented in Fig.~\ref{Figure2} for $\tilde{\alpha}/J$ $=0$, and the doping-independent effective amplitudes $\tilde{\alpha}/J=$ $0.1$ and $0.25$. 
	}
 }
\end{figure}

\begin{figure*}[tbp]%
	\includegraphics[width=0.33\textwidth]{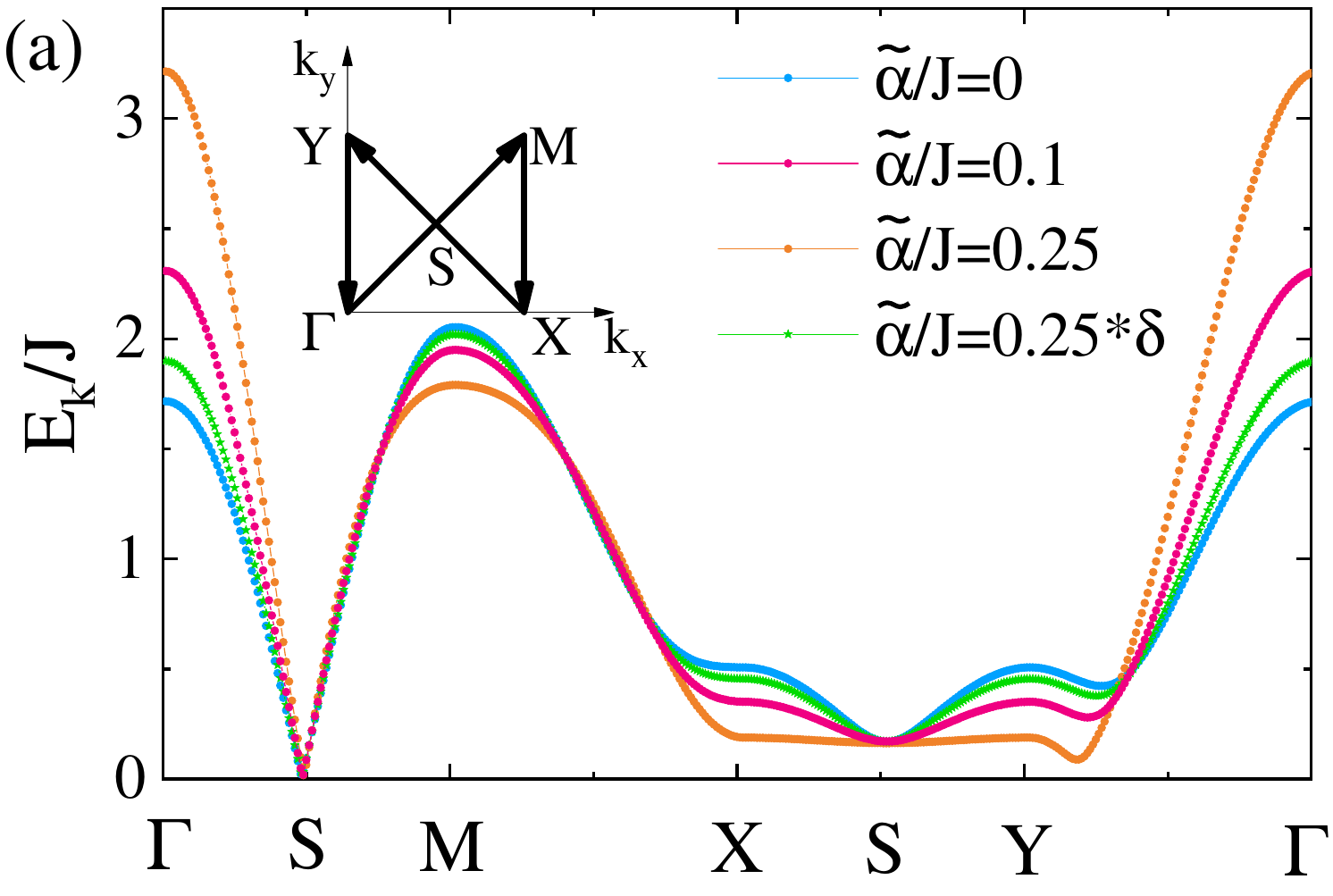}	
 \includegraphics[width=0.33\textwidth]{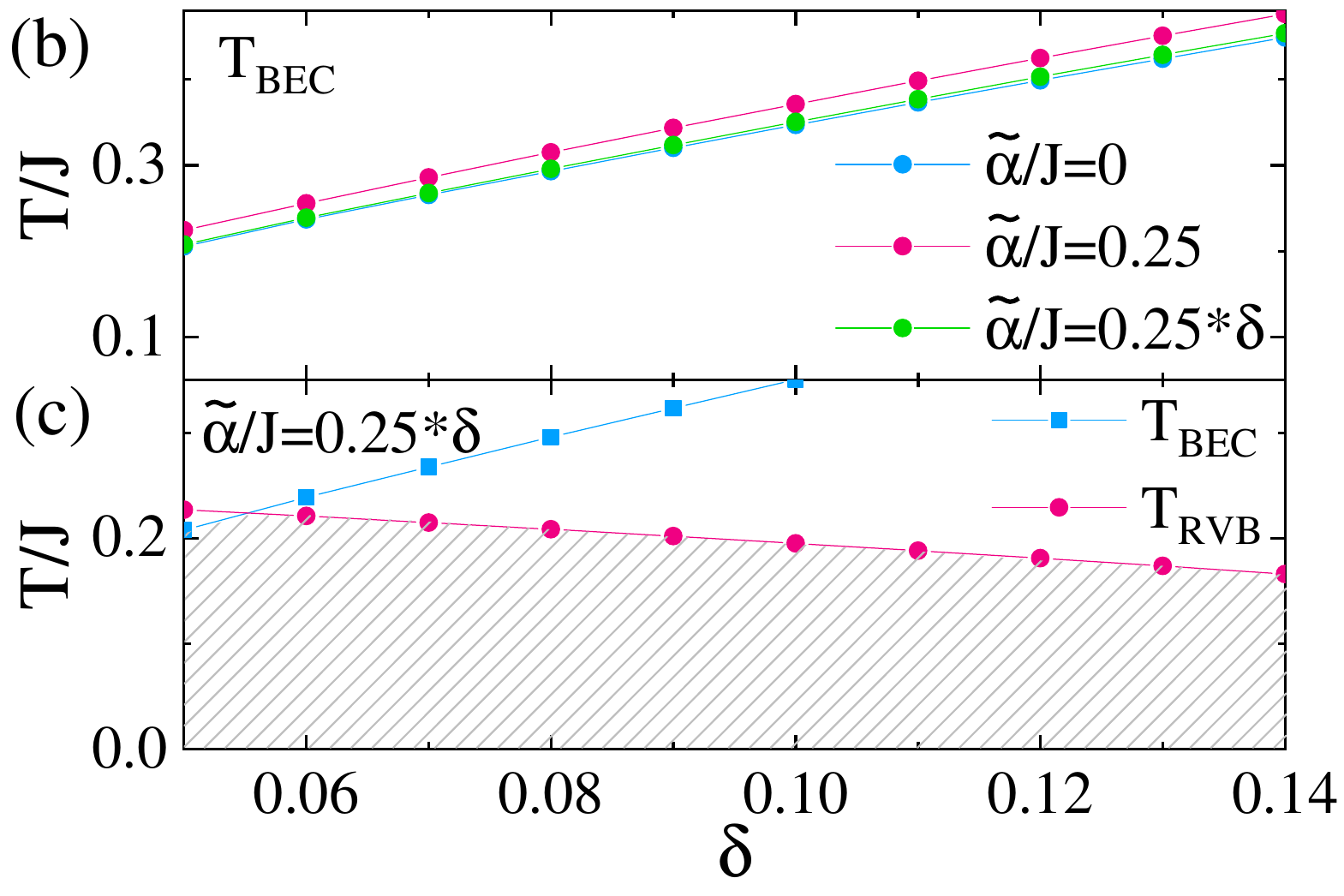}	
	\includegraphics[width=0.33\textwidth]{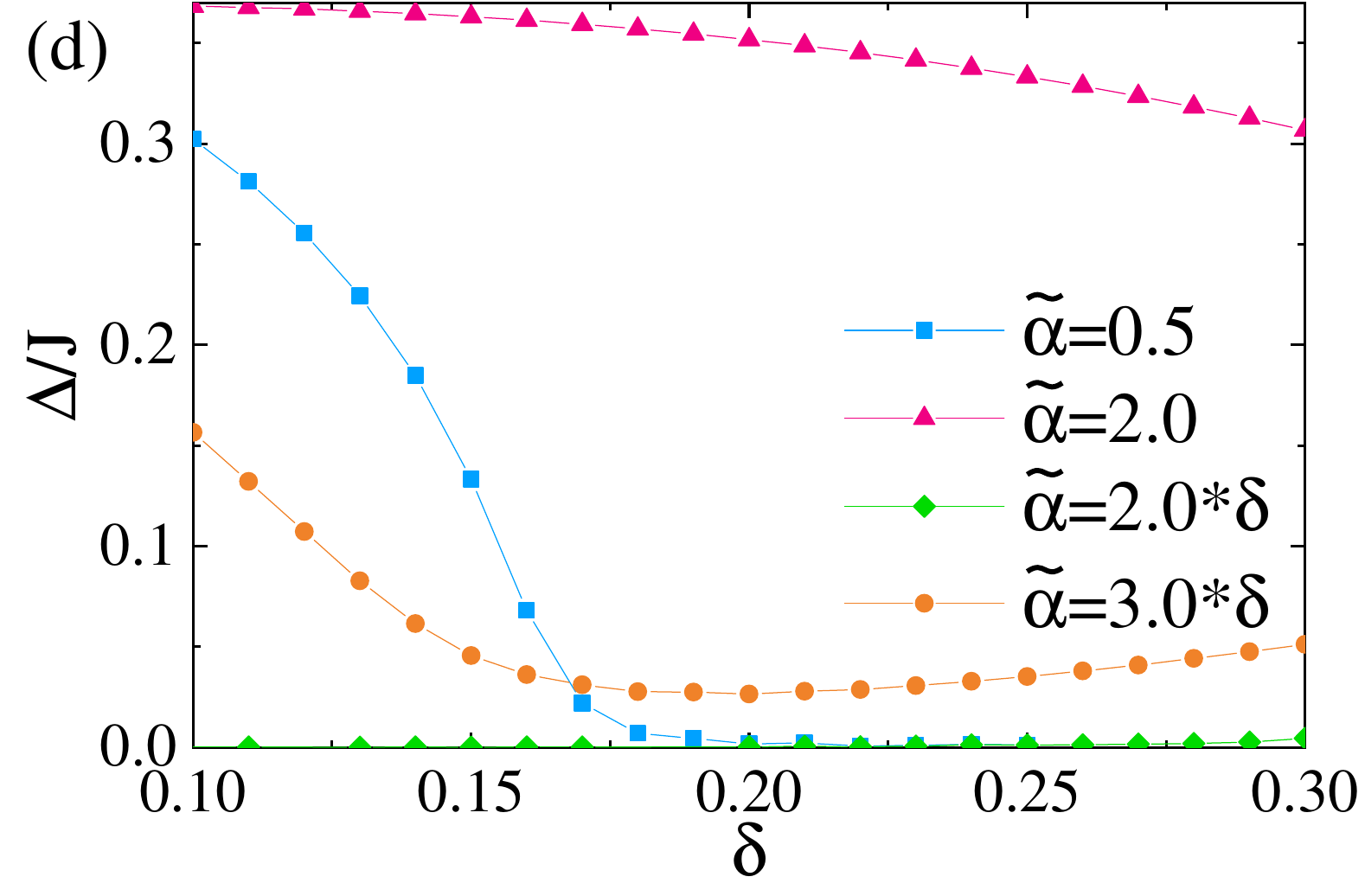}		
 
	\caption{\label{Figure9}(Color online)
{
  Quasiparticle dispersion $E_{\mathbf k}$,  boson-condensation temperature $T_{\mathrm{BEC}}$, and $s$-wave solutions for large $\alpha$, with the effective three-site hopping amplitude $\tilde{\alpha}$.
   Here, we consider $t/J=3$. In panels (a,b), we compare results using $\tilde{\alpha}/J=0.25\delta$  with those obtained using $\tilde{\alpha}/J=0,0.1,0.25$ which is presented in Fig.~\ref{Figure3}(d) and Fig.~\ref{Figure4}. Panel (c) shows the superconducting regime for $\tilde{\alpha}/J=0.25\delta$, indicated by the shaded area.  In panel (d), results for large $\alpha$ using $\tilde{\alpha}=\delta\alpha$  are compared with those using $\tilde{\alpha}=\alpha$ which is presented in Fig.~\ref{Figure7}. The quasiparticle dispersion $E_{\mathbf k}$ at doping concentration $\delta=0.125$ is plotted along the high symmetry line, with the path illustrated in the inset. The critical temperature $T_{\mathrm{BEC}}$ is calculated with $r_z=0.01$. 
	}
 }
\end{figure*}

\section{Results for the doping-dependent three-site hopping amplitude in SBMF solution.}\label{SecA1}
{
The three-site hopping term defined in  Hamiltonian~\eqref{H3s} can be rewritten in the SBMF description with an effective three-site hopping amplitude $\tilde{\alpha}$, as follows:
\begin{align}
&&\mathcal{H}_{\mathrm{\alpha}}=-\tilde{\alpha}\sum^{{\bf i}\ne {\bf k}}_{\langle \bf{i,j,k}\rangle,\sigma}(f^{\dag}_{{\bf i}\sigma}f^{\dag}_{{\bf j}\bar{\sigma}}f_{{\bf j}\bar{\sigma}}f_{{\bf k}\sigma}-f^{\dag}_{{\bf i}\sigma}f^{\dag}_{{\bf j}\bar{\sigma}}f_{{\bf j}\sigma}f_{{\bf k}\bar{\sigma}}).
\end{align}
where the effective amplitude $\tilde{\alpha}=\alpha\delta$ accounts for the doping-dependent three-site hopping amplitude, and when  $\tilde{\alpha}=\alpha$  is used, it is equivalent to the doping-independent three-site hopping amplitude employed in the main text. In this section, we evaluate the SBMF solutions that use the doping-dependent effective amplitude $\tilde{\alpha}/J=0.25\delta$ and compare them with $\tilde{\alpha}/J$ $=0$ and the doping-independent effective amplitudes $\tilde{\alpha}/J=$ $0.1$ and $0.25$, which are presented in the main text. }

{
In Fig.~\ref{Figure8}, we estimate the suppression of $d$-wave superconductivity induced by doping-dependent effective amplitude $\tilde{\alpha}/J=0.25\delta$, and compare these results with those presented in Fig.\ref{Figure2}.  The $d$-wave superconductivity of $\tilde{\alpha}/J=0.25\delta$ is close to that of $\tilde{\alpha}/J=0$ at low doping levels, and the suppression induced by $\tilde{\alpha}/J=0.25\delta$ becomes stronger as $\delta$ increases. At higher doping levels, the suppressive effect is close to that of the doping-independent effective amplitude $\tilde{\alpha}/J=0.1$. 
In Fig.~\ref{Figure9}, we examine other properties within the mean-field framework, including quasiparticle dispersion $E_{\mathbf k}$, boson-condensation temperature $T_{\mathrm{BEC}}$, and $s$-wave solutions.  As shown in Figs.~\ref{Figure9} (a-c), the modification of quasiparticle dispersion $E_{\mathbf k}$ and the impact on the condensation temperature caused by using the doping-dependent effective amplitude align with those from the main text.  The phase diagram is still primarily influenced by RVB pairing. 
To search for $s$-wave solutions, a larger $\alpha$ is required with the doping-dependent effective amplitude than with a doping-independent effective amplitude, as shown in Fig.~\ref{Figure9} (d). We find that using a doping-dependent effective amplitude still enhances $s$-wave. Then a larger $\alpha$ at intermediate doping levels will produce an effective $\tilde{\alpha}$ sufficient to support $s$-wave solutions.  As displayed in Fig.~\ref{Figure9} (d) for $\tilde{\alpha}/J=3\delta$, an increase in $\Delta$ is observed as doping increases at intermediate doping levels. This increase is attributed to the doping-dependent $\tilde{\alpha}$ and the relatively stable $\Delta$ for large $\alpha$ in the intermediate doping region as shown in Fig.~\ref{Figure7}.  More discussions on the $s$-wave solutions can be found in Appendix~\ref{SecA3}. 
In Fig.~\ref{Figure10}, the impacts of system sizes and aspect ratio of the square lattice on SC order are explored using the doping-dependent effective amplitude. The result with a doping-dependent $\tilde{\alpha}$ is similar to that of $\tilde{\alpha}=0$.  However, the differences increase with a larger $\delta$. 
In summary, these results, which use a doping-dependent $\tilde{\alpha}$, provide conclusions that are qualitatively consistent with those obtained using a doping-independent $\tilde{\alpha}$ employed in the main text, but show a weakened impact.
}

\section{Slave-boson mean-field approximation}\label{SecA2}
The resonating valence bond (RVB) theory successfully understands the superconducting mechanism in cuprates from a strong-correlation viewpoint~\cite{SingleH1,RVB1,RVB1.2}. The ground state is approximated by a Gutzwiller projected BCS-like wave function. The spins are assumed to form spin-singlet valence bonds in pairs. Superconductivity is induced when the half-filled Mott insulator is doped with electrons or holes. Under the slave-boson description, the SC order is characterized by the condensation of holons and RVB pairing of spinons.

{ We have the electron annihilation operator $c_{\mathbf{i}\sigma}= b^\dag_{\mathbf{i}}f_{\mathbf{i}\sigma}$}. And under the local constraint, we have $b_{\mathbf{i}}^{\dag}b_{\mathbf{i}}+\sum_{\sigma}f^{\dag}_{\mathbf{i}\sigma}f_{\mathbf{i}\sigma}=1$. This decomposition  will reproduce the Hamiltonian \eqref{Model} into a slave-boson representation. The charge hopping term can be written as:
\begin{align}
&&\mathcal{H}_{\mathrm{t}}=-\sum_{\langle \mathbf {i},\mathbf{j}\rangle}\sum_{\sigma}t(b_{\mathbf j}^\dag b_{\mathbf i} f^\dag_{\mathbf{i}\sigma}f_{\mathbf{j}\sigma}+b_{\mathbf i}^\dag b_{\mathbf j} f^\dag_{\mathbf{j}\sigma}f_{\mathbf{i}\sigma}).
\end{align}
By introducing the bond order parameter for spinons $\chi_{\bf{ij}}=\langle \sum_{\sigma}f^{\dag}_{\bf{i}\sigma}f_{\bf{j}\sigma}\rangle\equiv \chi$ and assuming the bond order parameter for holons $\langle b^{\dag}_{\bf{i}}b_{\bf{j}}\rangle\equiv \delta$, we can get a mean-field decoupling  as:
\begin{align}
	\mathcal{H}^{\mathrm{MF}}_{\mathrm{t}}=-t\sum_{\langle \bf i,j\rangle}	\chi_{\bf ij}(b_{\bf j}^\dag b_{\bf i}+ \! \text{H.c.} )+\delta\sum_{\sigma}(f^{\dag}_{\bf{i}\sigma}f_{\bf{j}\sigma}+ \! \text{H.c.}).
\end{align}
As a result of approximation $(1+\delta)^2\simeq 1$,  the spin exchange and density exchange terms can be written only in terms of
spinon operators, given by
\begin{eqnarray}
&&	{\bf{S}}_{\bf i}\cdot{\bf{S}}_{\bf j}=\frac{1}{4}\sum_{\sigma}2f^{\dag}_{{\bf{i}}\sigma}f_{{\bf{i}}\bar{\sigma}}f^{\dag}_{{\bf{j}}\bar{\sigma}}f_{{\bf{j}}\sigma}+f^{\dag}_{{\bf{i}}\sigma}f_{{\bf{i}}\sigma}f^{\dag}_{{\bf{j}}\sigma}f_{{\bf{j}}\sigma}-f^{\dag}_{{\bf{i}}\sigma}f_{{\bf{i}}\sigma}f^{\dag}_{{\bf{j}}\bar{\sigma}}f_{{\bf{j}}\bar{\sigma}},\nonumber\\
&&	n_{\bf i}n_{\bf j}=\sum_{\sigma}f^{\dag}_{{\bf{i}}\sigma}f_{{\bf{i}}\sigma}f^{\dag}_{{\bf{j}}\sigma}f_{{\bf{j}}\sigma}+f^{\dag}_{{\bf{i}}\sigma}f_{{\bf{i}}\sigma}f^{\dag}_{{\bf{j}}\bar{\sigma}}f_{{\bf{j}}\bar{\sigma}}.
\end{eqnarray}
 Then we can get the exchange term written in the slave-boson description, reads
\begin{align}
&&\mathcal{H}_{\mathrm{J}}=\frac{1}{2}J\sum_{\langle {\bf  i,j}\rangle,\sigma}f^{\dag}_{{\bf{i}}\sigma}f_{{\bf{i}}\bar{\sigma}}f^{\dag}_{{\bf{j}}\bar{\sigma}}f_{{\bf{j}}\sigma}-f^{\dag}_{{\bf{i}}\sigma}f_{{\bf{i}}\sigma}f^{\dag}_{{\bf{j}}\bar{\sigma}}f_{{\bf{j}}\bar{\sigma}}.
\end{align}
By introducing the  RVB pairing order parameter $\Delta _{\bf{ij}}=\langle f_{\bf{i}\downarrow}f_{\bf{j}\uparrow}-f_{\bf{i}\uparrow}f_{\bf{j}\downarrow}\rangle\equiv\Delta_{ x(y)}$ with $d$-wave pairing symmetry, i.e., $\Delta_{ x}=-\Delta_{ y}=\Delta$. Here, $\Delta_{ x(y)}$ represents NN bonds along $x(y)$-direction.  The mean-field Hamiltonian takes the form
 \begin{eqnarray}
	\mathcal{H}^{\mathrm{MF}}_{\mathrm{J}}=&&-\sum_{\langle {\bf i,j}\rangle}[\frac{1}{2}J\big(\Delta^*_{\bf ij}(f_{{\bf i}\downarrow}f_{{\bf j}\uparrow}-f_{{\bf i}\uparrow}f_{{\bf j}\downarrow})+ \! \text{H.c.}\big)\nonumber\\
	&&+\frac{1}{4}J\chi_{\bf ij}\sum_{\sigma}(f^{\dag}_{{\bf{i}}\sigma}f_{{\bf{j}}\sigma}+\! \text{H.c.})\nonumber\\
	&&+\frac{1}{4}J(1-\delta)\sum_{\sigma}(  f^{\dag}_{{\bf{i}}\sigma}f_{{\bf{i}}\sigma}+ f^{\dag}_{{\bf{j}}\sigma}f_{{\bf{j}}\sigma})].
\end{eqnarray}
The three-site hopping term defined in  Hamiltonian~\eqref{H3s} can be similarly rewritten {at the same order as the Heisenberg term when  neglecting the doping dependence}, as follows
\begin{align}
&&\mathcal{H}_{\mathrm{\alpha}}=-\alpha\sum^{{\bf i}\ne {\bf k}}_{\langle \bf{i,j,k}\rangle,\sigma}(f^{\dag}_{{\bf i}\sigma}f^{\dag}_{{\bf j}\bar{\sigma}}f_{{\bf j}\bar{\sigma}}f_{{\bf k}\sigma}-f^{\dag}_{{\bf i}\sigma}f^{\dag}_{{\bf j}\bar{\sigma}}f_{{\bf j}\sigma}f_{{\bf k}\bar{\sigma}}).
\end{align}
We decouple the four-spinon term into particle-particle channel and particle-hole channel,  the mean-field Hamiltonian reads

 \begin{eqnarray}
\mathcal{H}_{\mathrm{\alpha}}^{\mathrm{MF}}=&&-\alpha\sum^{{\bf i}\ne {\bf k}}_{\langle \bf{i,j,k}\rangle}[\big(\Delta^*_{\bf ij}(f_{{\bf j}\downarrow}f_{{\bf k}\uparrow}-f_{{\bf j}\uparrow}f_{{\bf k}\downarrow})\nonumber\\
&&+\Delta_{\bf jk}(f^\dag_{{\bf i}\uparrow}f^\dag_{{\bf j}\downarrow}-f^\dag_{{\bf i}\downarrow}f^\dag_{{\bf j}\uparrow})\big)\nonumber\\
&&+\frac{1}{2}	\sum_{\sigma}\big(\chi_{\bf ik}f^\dag_{{\bf j}\sigma}f_{{\bf j}\sigma}+(1-\delta) f^\dag_{{\bf i}\sigma}f_{{\bf k}\sigma}\nonumber\\	
&&+\chi_{\bf ij}f^\dag_{{\bf j}\sigma}f_{{\bf k}\sigma}+\chi_{\bf jk}f^\dag_{{\bf i}\sigma}f_{{\bf j}\sigma}\big)].
\end{eqnarray}

\begin{figure*}[tbp]%

		\includegraphics[width=0.48\textwidth]{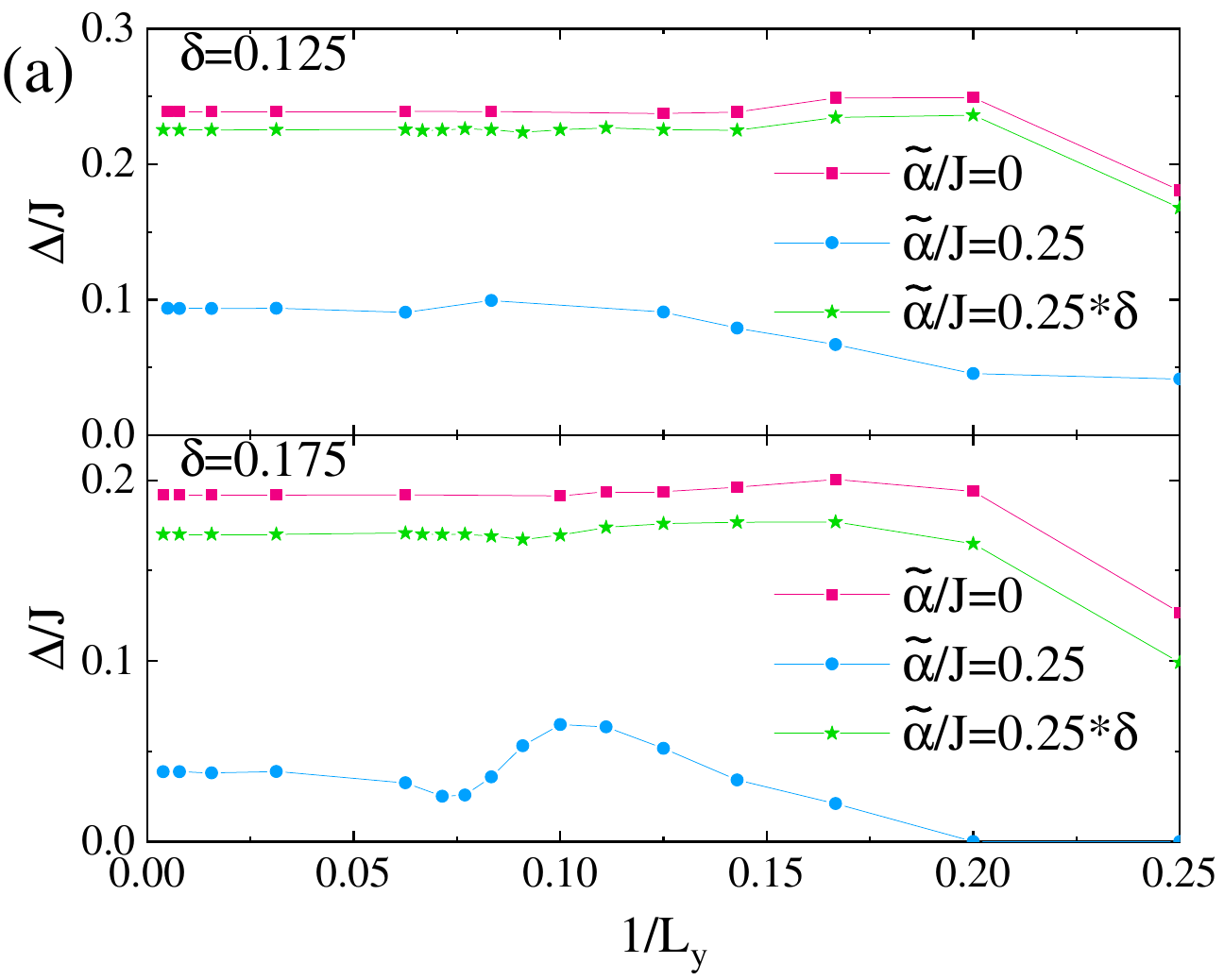}	
		\includegraphics[width=0.48\textwidth]{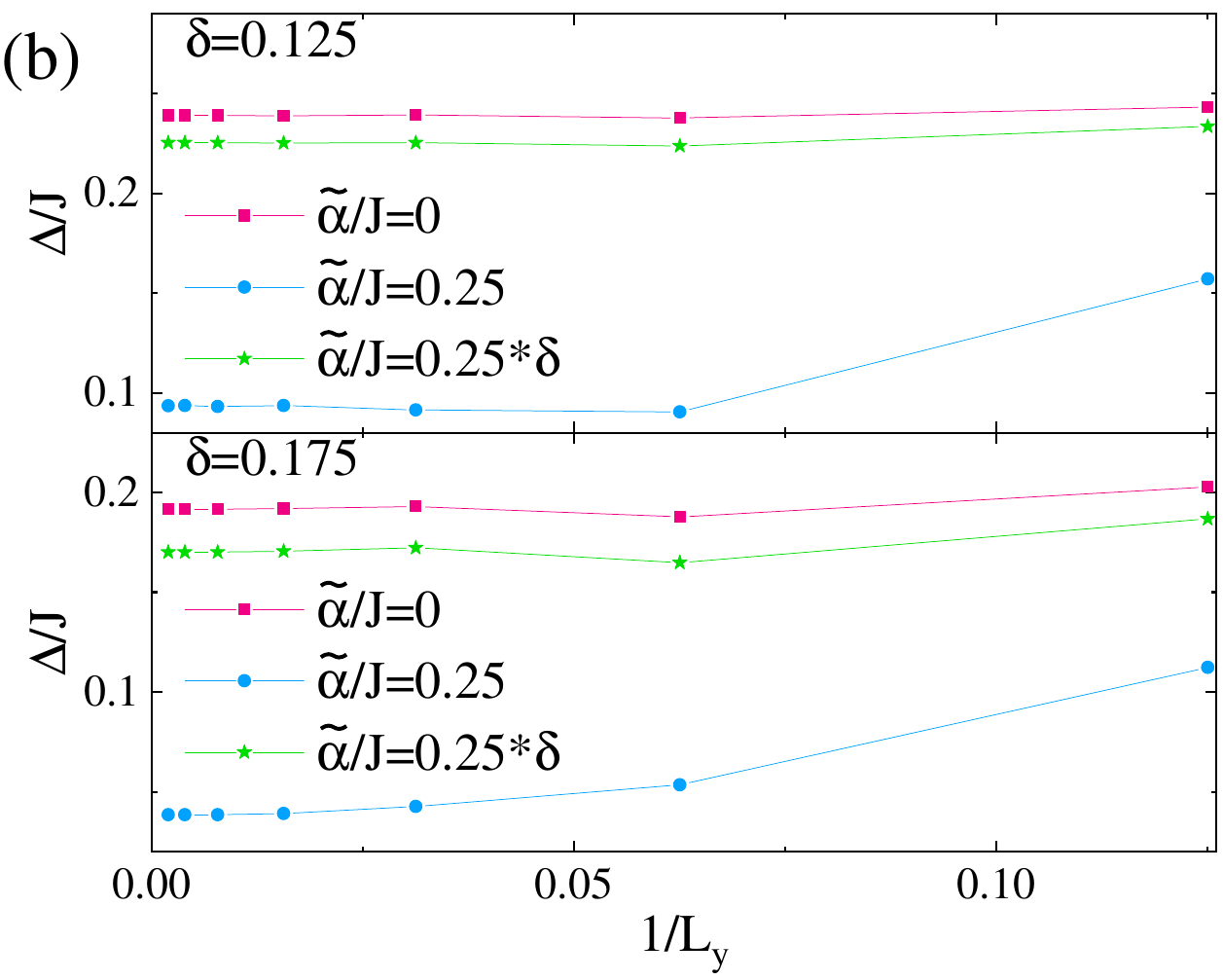}	
\caption{\label{Figure10}(Color online)
		 {
   {
   The superconducting order parameter $\Delta$ versus the inverse of width $L_y$ using the doping-dependent effective three-site hopping amplitude $\tilde{\alpha}/J=0.25\delta$. Here, we compare the results with those presented in Fig.~\ref{Figure6}  for $\tilde{\alpha}/J$ $=0$, and the doping-independent effective amplitude $\tilde{\alpha}/J=$ $0.25$. The doping concentrations are set at $\delta=0.125$ and $\delta=0.175$ with a fixed ratio $t/J=3$.  In panel (a), we fix $L_x=200$  and vary aspect ratio $r$ down to $1$.  In panels (b), we fix $r=1$ and vary system size $N$ up to $200\times200$.}
	}
 }
\end{figure*}

The SBMF Hamiltonian for $t$-$J$-$\alpha$ model  is written as $\mathcal{H}_{\mathrm{SBMF}}=\mathcal{H}^{\mathrm{MF}}_{\mathrm{t}}+\mathcal{H}^{\mathrm{MF}}_{\mathrm{J}}+\mathcal{H}^{\mathrm{MF}}_{\mathrm{\alpha}}$.  While the  $\mathcal{H}_{0}$ term  includes  constant contributions, reads
\begin{eqnarray}\label{h0}
	&&\mathcal{H}_{0}/N=(J-2\alpha) \Delta^2+(\frac{1}{2}J+3\alpha)\chi^2\nonumber\\
	&&+4t\delta\chi+\frac{1}{2}J(1-\delta)^2-(\mu_f+\mu_b)\delta+\mu_f.\label{H01}
\end{eqnarray}	

The chemical potential for spinons $\mu_f$ includes Lagrange multiplier constant $\lambda$ and chemical potential $\mu$. The chemical potential for holons is $\mu_b=\mu$.  They are determined from self-consistent equations [see Ref.~\cite{Review2,SBMF2}]
\begin{eqnarray}
	&&1-\delta=\frac{1}{N}\sum_{\bf{k}}[1-\frac{\epsilon_{\bf{k}}}{E_{\bf{k}}}\tanh(\frac{\beta E_{\bf{k}}}{2})],\label{nufse}\\
	&&\delta=\frac{1}{N}\sum_{\bf{k}} \frac{1}{e^{\beta\omega_{\bf{k}}}-1}.	\label{mufse}
\end{eqnarray}

\begin{figure*}[tbp]%
	\includegraphics[width=0.48\textwidth]{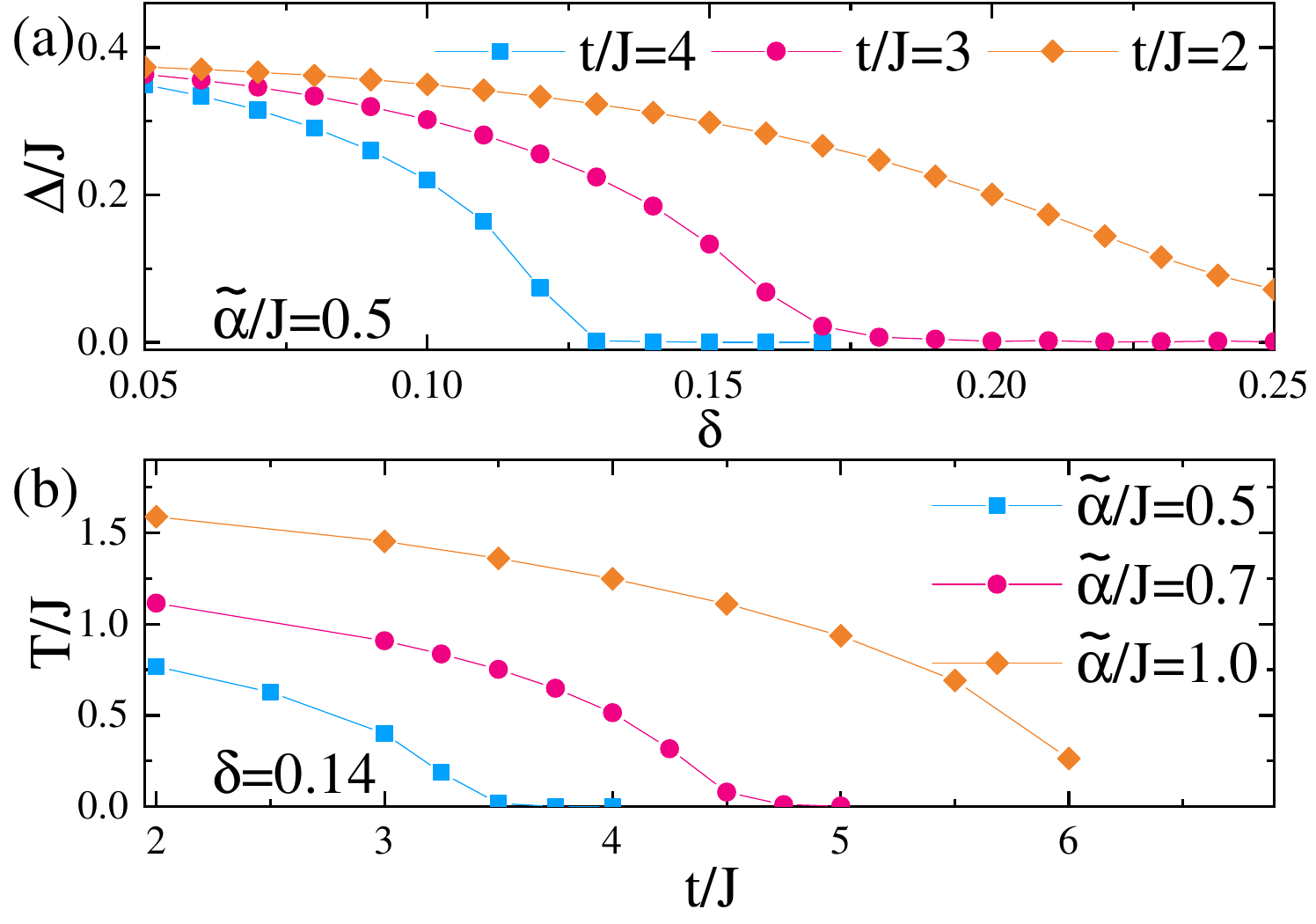}				
	\includegraphics[width=0.48\textwidth]{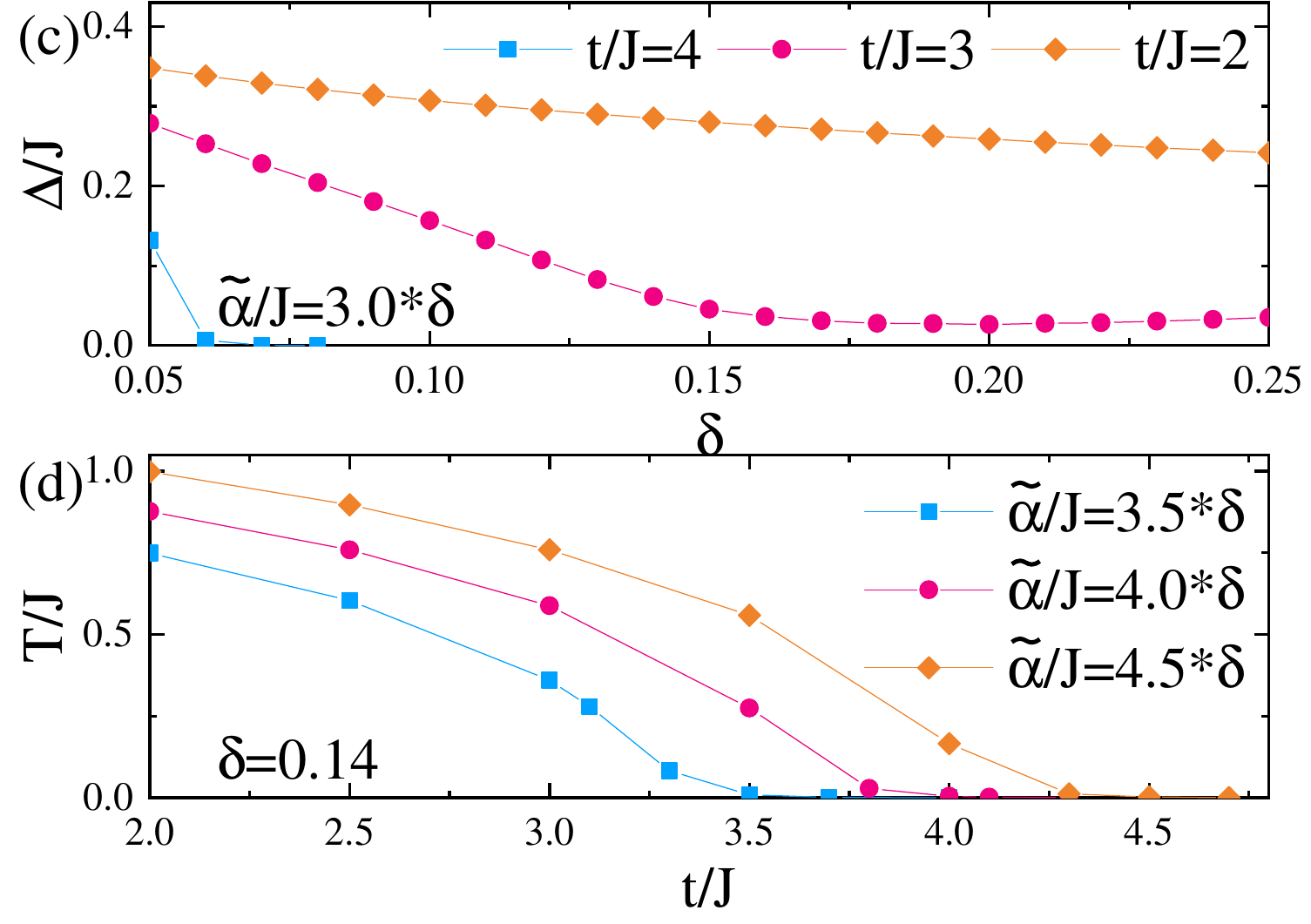}		
 
	\caption{\label{Figure11}(Color online)
 {
   The impact of coupling strength on $s$-wave solution. The results are presented for using both doping-independent effective amplitude $\tilde{\alpha}=\alpha$ [panels (a) and (b)]
and doping-dependent effective amplitude $\tilde{\alpha}=\delta\alpha$ [panels (c) and (d)].  In panels (a) and (c), the $s$-wave RVB pairing order parameter $\Delta$ is plotted against doping concentration $\delta$. Here, we consider $t/J=2,3,4$ and $T=0$. We fix $\tilde{\alpha}/J=0.5$ in panel (a)  and $\tilde{\alpha}/J=3\delta$ in panel (c).  Increasing coupling strength $t/J$ intensely suppresses the $s$-wave pairing.
    In panels (b) and (d), the critical temperature $T_{\mathrm{RVB}}$ is plotted against coupling strength $t/J$. Here, we fix doping concentration $\delta=0.14$. We consider $\tilde{\alpha}/J=0.5,0.7,1.0$ in panel (b) and  $\tilde{\alpha}/J=3.5\delta,4.0\delta,4.5\delta$ in panel (d). The $T_{\mathrm{RVB}}$ curve for $s$-wave solution exhibits a sharp drop in a broad range of increasing $t/J$.
     }
     }
\end{figure*}

%%%%%%%%%%%%%%%%%%%%%%%%%%%%%%%%%%%%%%%%%%%%%%%%%%%%%%
\begin{figure}[tbp]
\begin{center}
\includegraphics[width=0.48\textwidth]{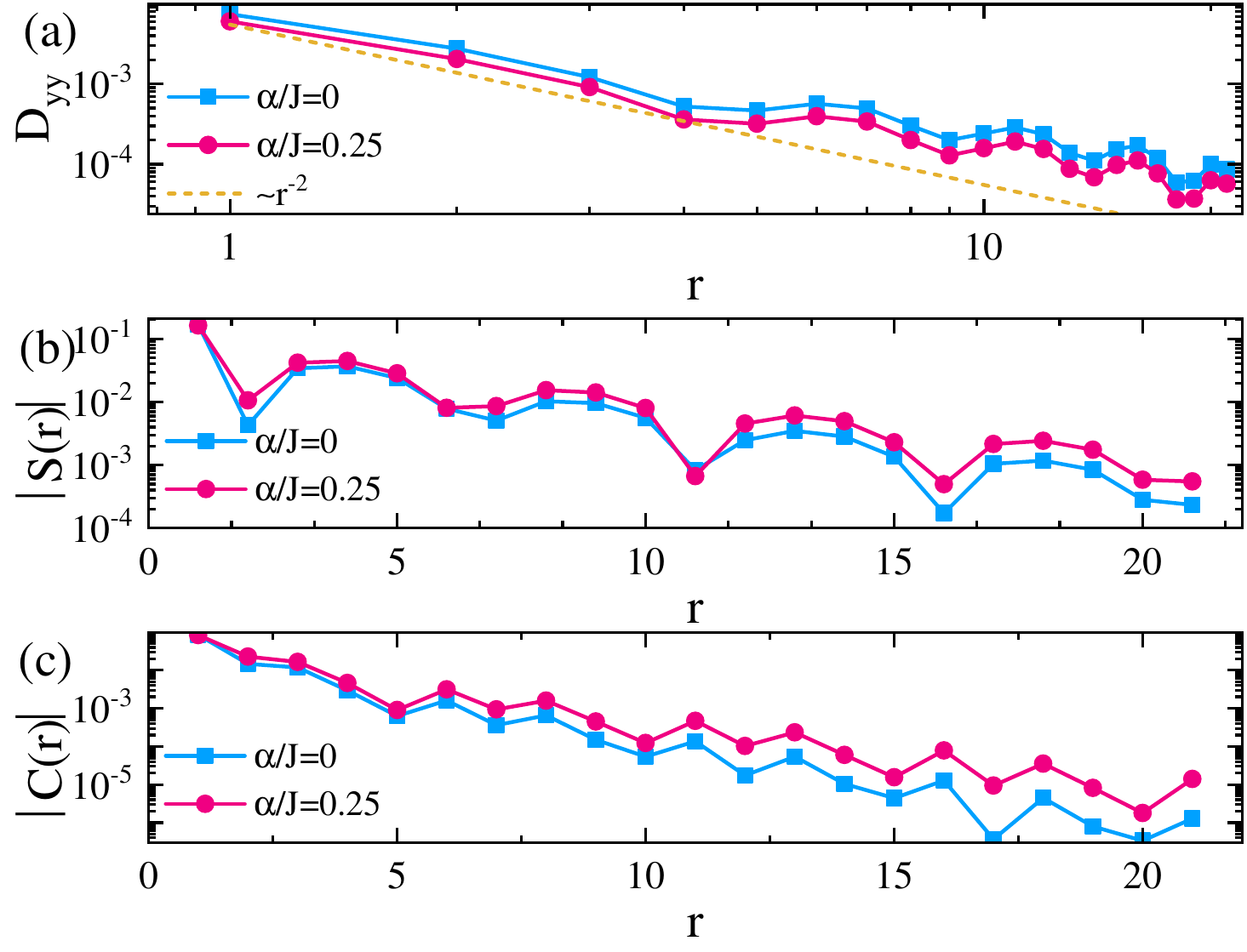}
\end{center}
\par
\caption{{Correlation functions from DMRG simulation}.
(a) Double-logarithmic plot of the pairing correlations $D_{yy}(r)$.
(b) Semi-logarithmic plot of the spin correlations $|S(r)|$.
(c) Semi-logarithmic plot of the single particle propagator $|C(r)|$.
Here, we set $\delta=0.1$, $t/J=3$, $N=30\times 4$.
}
\label{Figure12}
\end{figure}
%%%%%%%%%%%%%%%%%%%%%%%%%%%%%%%%%%%%%%%%%%%%%%%%%%%%%%

 The mean-field Hamiltonian $\mathcal{H}_{\mathrm{SBMF}}$ in momentum space,   written in Eq.~\eqref{HSBMF}, can be expressed in a matrix form,
\begin{align}
&&	\mathcal{H}_{\mathrm{SBMF}}=\sum_{\bf{k}}	\frac{1}{2}\psi^{\dag}_{\bf k}M_{\bf k}\psi_{\bf k}+\sum_{\bf{k}} \omega_{\bf{k}} b^\dag_{\bf{k}}b_{\bf{k}} +\sum_{\bf{k}}\epsilon_{\bf k},
\end{align}
where $\psi_{\bf k}=(f_{{\bf k}\downarrow}, f_{{\bf k}\downarrow}^\dag, f_{-{\bf k}\uparrow}, f_{-{\bf k}\uparrow}^\dag)^T$ and
\begin{align}
M_{\bf k}=	\begin{pmatrix}
	\epsilon_{\bf k}&0&0&\Delta_{\bf k}\\
	0&-\epsilon_{\bf k}&-\Delta^*_{\bf k}&0\\		
	0&-\Delta_{\bf k}&\epsilon_{-\bf k}&0\\	
	\Delta^*_{\bf k}&0&0&-\epsilon_{-\bf k}\\		
\end{pmatrix}.
\end{align}
By applying $\psi_{\bf k}$ to  $\Psi_{\bf k}=(\alpha_{\bf k}, \alpha_{\bf k}^\dag, \beta_{\bf k}, \beta_{\bf k}^\dag)^T$  with a Bogoliubov transformation, we can get the diagonal Hamiltonian
\begin{align}
\mathcal{H}_{\mathrm{SBMF}}=\sum_{\bf{k}}\frac{1}{2}\Psi^{\dag}_{\bf k}\Lambda_{\bf k}\Psi_{\bf k}+\sum_{\bf{k}} \omega_{\bf{k}} b^\dag_{\bf{k}}b_{\bf{k}}+\sum_{\bf{k}}\epsilon_{\bf k}.
\end{align}
Where $E_{\bf{k}}=\sqrt{\vert\Delta_{\bf{k}}\vert^2+\epsilon_{\bf{k}}^2}$ and
\begin{align}
\Lambda_{\bf k}=	\begin{pmatrix}
	E_{\bf k}&0&0&0\\
	0&-E_{\bf k}&0&0\\		
	0&0&E_{-\bf k}&0\\	
	0&0&0&-E_{-\bf k}\\	
\end{pmatrix}	
\end{align}
The order parameters can be self-consistently determined by minimizing the free energy written in Eq.~\eqref{Free}.

\section{The  $s$-wave solution.}\label{SecA3}

The $s$-wave pairing and $d$-wave pairing give rise to BCS gap functions such as $\cos k_x+\cos k_y$ and $\cos k_x-\cos k_y$,  respectively~\cite{3s8}. We numerically search for solutions that minimize the Ginzburg-Landau free energy within the mean-field approximation, indicating both $s$-wave and $d$-wave solutions to be theoretically viable in self-consistent calculations. In the pure $t$-$J$ model, the $d$-wave pairing is generally favored, while the three-site hopping term enhances the $s$-wave component~\cite{3s8, 3s11, ap201} [{see our results in Fig.~\ref{Figure7} and Fig.~\ref{Figure9}(d)}]. {In this section, results are presented for using both doping-independent effective amplitude, i.e., $\tilde{\alpha}=\alpha$, and doping-dependent effective amplitude, i.e., $\tilde{\alpha}=\delta\alpha$ [see Appendix~\ref{SecA1}].}

We investigate the impact of coupling strength on $s$-wave solution in Fig.~\ref{Figure11}.  In panels (a,c), the results demonstrate the intense suppression in the $t$-$J$-$\alpha$ model, in agreement with earlier reports~\cite{ap201}. In panel (b,d), the curve of $s$-wave solution experiences a sharp drop in a broad range of $t/J\approx 2\sim 6$. By contrast, the results of $d$-wave solution shown in {Fig.\ref{Figure8} (c) } exhibit a slight linear decrease.

\section{Correlation functions}\label{SecA4}

For the $t$-$J$ model with three-site hopping term, we analyze  the pair correlations $D_{yy}({r})$, the spin correlations $S(r)$, and the single particle propagator $C(r)$, which are defined as
\begin{equation}\label{eqS:D}
\begin{split}
D_{\alpha\beta}({r}) &\equiv \left\langle \hat{\Delta}^{\dagger}_{\alpha}(\mathbf{r}_0) \hat{\Delta}_{\beta}(\mathbf{r}_0 +{r}\boldsymbol{e}_x)\right\rangle,\\
S(r)&\equiv \langle \mathbf{S}_{{\mathbf{r}}_0}\cdot \mathbf{S}_{{\mathbf{r}_0+r\boldsymbol{e}_x}}\rangle,\\
C(r)&\equiv\sum_{\sigma}\langle c_{\mathbf{r}_0,\sigma}^\dagger c_{{\mathbf{r}_0+r\boldsymbol{e}_x},\sigma}\rangle .
\end{split}
\end{equation}
Here, $\hat{\Delta}_{\alpha}^\dagger(\mathbf{r}_0) \equiv \frac  1 {\sqrt{2}}(c_{\mathbf{r}_0+\boldsymbol{e}_{\alpha}, \downarrow}^{\dagger} c_{\mathbf{r}_0, \uparrow}^{\dagger}-c_{\mathbf{r}_0+\boldsymbol{e}_{\alpha}, \uparrow}^{\dagger} c_{\mathbf{r}_0, \downarrow}^{\dagger})$ and $\alpha, \beta$ denote the bonds $\alpha, \beta = x, y$.
Despite the inclusion of the three-site hopping, the qualitative behaviors of these correlations---power-law decay for $D_{yy}(r)$ and exponential decay for both $|S(r)|$ and $|C(r)|$---remain unchanged for 4-leg cylinder with $N=30\times 4$, $\delta=0.1$ and $t/J=3$, as illustrated in Fig.~\ref{Figure12}. This finding underscores the robustness of correlation behavior in the presence of the three-site term.

\end{document}